\documentclass[aps,prd,groupaddress,tighten,twocolumn,nofootinbib,superscriptaddress]{revtex4}
\usepackage{amssymb}
\usepackage{amsmath,color}
\usepackage{epsfig}
\usepackage{graphicx,epsf,epsfig}
\usepackage{bbm}
\def\slc#1{\setbox0=\hbox{$#1$}           % set a box for #1
    \dimen0=\wd0                                 % and get its size
    \setbox1=\hbox{/} \dimen1=\wd1               % get size of /
    \ifdim\dimen0>\dimen1                        % #1 is bigger
       \rlap{\hbox to \dimen0{\hfil/\hfil}}      % so center / in box
       #1                                        % and print #1
    \else                                        % / is bigger
       \rlap{\hbox to \dimen1{\hfil$#1$\hfil}}   % so center #1
       /                                         % and print /
    \fi}

\begin{document}
%----------------------------------------------------------------------------------
\title{Higgs triplets at like-sign linear colliders and neutrino mixing
}
%----------------------------------------------------------------------------------
%\date{\today}
%----------------------------------------------------------------------------------
\author{Werner Rodejohann}
\email{werner.rodejohann@mpi-hd.mpg.de}

\affiliation{Max-Planck-Institut f{\"u}r Kernphysik, Postfach
103980, 69029 Heidelberg, Germany}

\author{He Zhang}
\email{he.zhang@mpi-hd.mpg.de}

\affiliation{Max-Planck-Institut f{\"u}r Kernphysik, Postfach
103980, 69029 Heidelberg, Germany}

%----------------------------------------------------------------------------------
%\pacs{12.10.-g, 12.60.Jv, 14.60.Pq, 12.15.Ff}
%----------------------------------------------------------------------------------

\begin{abstract}
We study the phenomenology of the type-II seesaw model at a linear
$e^-e^-$ collider. We show that the process $e^-e^- \rightarrow
\alpha^- \beta^-$ ($\alpha, \beta = e, \mu, \tau$ being charged
leptons) mediated by a doubly charged scalar is very sensitive to
the neutrino parameters, in particular the absolute neutrino mass
scale and the Majorana CP-violating phases. We identify the regions
in parameter space in which appreciable collider signatures in the
channel with two like-sign muons in the final state are possible.
This includes Higgs triplet masses beyond the reach of the LHC.
\end{abstract}
\maketitle
%%%%%%%%%%%%%%%%%%%%%%%%%%%%%%%%%%%%%%%%%%%

\section{Introduction} \label{sec:intro}

The origin of neutrino masses and leptonic flavor mixing emerges as
one of the most challenging problems in particle physics. Among
various theories of this kind, the seesaw
mechanism~\cite{Minkowski:1977sc,Yanagida:1979as,GellMann:1980vs,Mohapatra:1979ia}
attracts a lot of attention in virtue of its naturalness and
simplicity in explaining the smallness of neutrino masses. In the
standard type-I seesaw model, the fermion sector of the Standard
Model (SM) is extended by adding right-handed neutrinos having large
Majorana masses $M_{\rm R}$. In its natural version the neutrino
masses are suppressed with respect to typical SM (Dirac) masses
$m_{\rm D}$ by a factor $m_{\rm D}/M_{\rm R}$. With $m_{\rm D}$ of
weak scale it follows that sub-eV neutrino masses require heavy
neutrino masses many orders of magnitude above the center-of-mass
energies of realistic colliders. In addition, as right-handed
neutrinos possess no gauge couplings, their production is suppressed
by a mixing factor of order $m_{\rm D}/M_{\rm R}$, so that all in
all the mechanism lacks testability. Only at the price of extreme
cancellations
\cite{Pilaftsis:1991ug,Kersten:2007vk,Zhang:2009ac,Adhikari:2010yt},
or by introducing additional gauge groups, one can achieve
production of type-I seesaw messengers at colliders.

In contrast, in the type-II seesaw
model~\cite{Schechter:1980gr,Lazarides:1980nt,Mohapatra:1980yp} one
extends the scalar sector of the SM by introducing an $SU(2)_{\rm
L}$ Higgs triplet, which couples to two lepton doublets and thereby
gives rise to a Majorana mass term of neutrinos after electroweak
symmetry breaking. This mass is given by a vacuum expectation value
times a Yukawa coupling. A Higgs triplet can be naturally embedded
in many frameworks, e.g., grand unified, left-right symmetric, or
little Higgs models. It is important to note that tiny neutrino mass
by no means requires that the mass of the Higgs triplet is huge. In
addition, Higgs triplets do possess gauge couplings, which
facilitates their production at colliders
\cite{Han:2007bk,Chao:2008mq,Perez:2008ha,delAguila:2008cj}. In such
a scenario, the bilepton decays of the doubly charged component is
firmly connected to the neutrino mass matrix, which opens a very
promising link between neutrino parameters and collider
signatures~\cite{Chun:2003ej,Akeroyd:2005gt,Garayoa:2007fw,Kadastik:2007yd,Akeroyd:2007zv,Chen:2008jh,Ren:2008yi,Petcov:2009zr,Nishiura:2009jn}.
Higgs triplets also induce lepton flavor violating (LFV) charged
lepton decays, which can be used to constrain the parameters
associated with them
\cite{Cuypers:1996ia,Frampton:1997gc,WR,Akeroyd:2009nu}. Moreover,
observable non-standard neutrino interaction effects induced by the
singly charged component of the Higgs triplet might be discovered in
future long-baseline neutrino oscillation
experiments~\cite{Malinsky:2008qn}.

The Large Hadron Collider is shown to be able to discover Higgs
triplets up to $600~{\rm GeV}\sim 1~{\rm TeV}$ depending on the
neutrino mass hierarchy~\cite{Perez:2008ha,delAguila:2008cj}.
Similarly, the doubly charged component of the Higgs triplet may
also be pair produced at a linear $e^+e^-$ collider via virtual
exchange of $Z^*$ and $\gamma^*$~\cite{Gunion:1989ci}. The decay of
the Higgs triplet in like-sign lepton pairs is then in analogy to
that at the LHC. In this paper, however, we will study the
production of doubly charged Higgs triplets at a linear collider in
the like-sign lepton mode. A linear $e^-e^-$ collider could provide
a substantial and complementary role in identifying new physics
beyond the SM. In such a running mode, a number of lepton number
violating (LNV) processes mediated by any LNV physics, including
Higgs triplets, can be explored to a very good precision, since they
are basically free from SM background. One $\Delta L=2$ process to
be investigated is the inverse neutrinoless double beta decay, i.e.,
$e^-e^- \rightarrow W^-W^-$, which is however suppressed by the
small vacuum expectation value of the Higgs triplet, and is not very
promising unless a very narrow resonance is
met~\cite{Rodejohann:2010jh}. In this work, we focus our attention
on the phenomena of bilepton production process mediated by a Higgs
triplet at a linear $e^-e^-$ collider:
\begin{eqnarray}
e^-  e^- \rightarrow \alpha^- \beta^- \, .
\end{eqnarray}
We demonstrate that the
collider signatures are firmly correlated to neutrino parameters,
while appreciable signals can be expected without the need of
resonant enhancement.
Such a process is impossible at the LHC, and proceeds via
$s$-channel exchange, see Fig.~\ref{fig:fig1}. The difference to the
aforementioned processes is the absence of gauge couplings and the
pure $s$-channel production of basically massless final state
particles; the cross section depends on Yukawa couplings (and hence
the neutrino flavor structure) only.

This work is organized as follows: In Sec.~\ref{sec:framework}, we
present the framework of the low-scale type-II seesaw model. In
Sec.~\ref{sec:collider signatures}, we focus on the characteristic
features of the processes mediated by the doubly charged scalar,
and summarize the current constraints on the relevant Yukawa
couplings. In particular, we figure out the intrinsic correlations
between neutrino parameters and the bilepton production processes at
a linear $e^-e^-$ collider. Numerical analysis and illustrations
will be performed in detail in Sec.~\ref{sec:numerics}. Finally, in
Sec.~\ref{sec:summary}, we summarize our results and conclude.

\section{Neutrino masses from the type-II seesaw} \label{sec:framework}

In the simplest type-II seesaw framework, one heavy Higgs triplet
with hypercharge $Y=2$ is introduced besides the SM particle
content. Apart from the SM interactions, one can further write down
a gauge invariant coupling between the Higgs triplet and two lepton
doublets as
\begin{eqnarray}
{\cal L}_\Delta = h_{\alpha \beta} \overline{L}_{\alpha} i\tau_2
\Delta L^c_\beta + {\rm H.c.},  \label{eq:L}
\end{eqnarray}
where $L_\alpha=(\nu_\alpha,\ell_\alpha)^T$ (for
$\alpha=e,\mu,\tau$) denote the lepton doublet, $h_{\alpha\beta}$ is
a symmetric Yukawa coupling matrix, and $\Delta$ is a $2\times2$
representation of the Higgs triplet
\begin{eqnarray}
\Delta = \left( \begin{matrix} \Delta^+/\sqrt{2} & \Delta^{++} \cr
\Delta^0 & -\Delta^+/\sqrt{2} \end{matrix} \right)  .
\end{eqnarray}
We further expand Eq.~\eqref{eq:L} and express the interactions in
terms of component fields, i.e.,
\begin{eqnarray}
{\cal L}_\Delta &=& h_{\alpha \beta} \left[ \Delta^0
\overline{\nu_\alpha} P_{\rm L} \nu^c_\beta -
\frac{1}{\sqrt{2}}\Delta^+ \left( \overline{\ell_\alpha} P_{\rm L}
\nu^c_\beta  + \overline{\nu_\alpha} P_{\rm L} \ell^c_\beta \right)
\right.  \nonumber \\ && \left.-\Delta^{++} \overline{\ell_\alpha}
P_{\rm L} \ell^c_{\beta}\right] + {\rm H.c.}
\end{eqnarray}
In the language of effective theory, the heavy Higgs triplet should
be integrated out from the full theory, and, at tree level, a
Majorana mass term of neutrinos is given by~\cite{Chao:2006ye}
\begin{eqnarray}
{\cal L}_\nu = h_{\alpha \beta} \frac{v_{\rm L}}{\sqrt{2}}
\overline{\nu_{\rm L \alpha}}\nu^c_{{\rm L}\beta}  + {\rm H.c.}  =
\frac{1}{2}m_{\alpha\beta}\overline{\nu_{\rm L \alpha}}\nu^c_{{\rm
L}\beta} + {\rm H.c.}, \label{eq:m}
\end{eqnarray}
where $v_{\rm L}$ is the vacuum expectation value (VEV) of the Higgs
triplet, i.e., $\langle \Delta^0\rangle = v_{\rm L}/\sqrt{2}$. The
main constraint on $v_{\rm L}$ stems from the electroweak $\rho$
parameter, and roughly gives $v_{\rm L} \lesssim 8~{\rm GeV}$. How
exactly $v_{\rm L}$ is related to the model parameters depends on
details of the underlying physics, it may be that $v_{\rm L} = v^2
\mu/m_{\Delta}^2 $, where $\mu$ is a dimensionful coupling of the
triplet with two Higgs doublets, or $v_{\rm L} \propto v^2/v_{\rm
R}$ in left-right symmetric theories, where $v_{\rm R}$ is the scale
of right-handed physics, i.e., the VEV of an $SU(2)_{\rm R}$
triplet. Here we will not speculate on the origin of the magnitude
of $m_\Delta$, but assume in a model-independent way that it is not
above TeV scale, and that the smallness of $v_{\rm L}$ is due to
other parameters in the underlying physics.

As usual, $m$ can be diagonalized by means of a unitary
transformation, viz.
\begin{eqnarray}\label{eq:U}
m  = U ~{\rm diag} (m_1,m_2,m_3) ~U^{T} \, ,
\end{eqnarray}
where $m_i$ ($i=1,2,3$) denote neutrino masses, and $U$ the leptonic
mixing matrix. In the standard (CKM-like) parametrization one
has
\begin{eqnarray}\label{eq:parametrization}
U  = R_{23}P_{\delta}R_{13}P_{\delta}^{-1}R_{12}P_{M} \, ,
\end{eqnarray}
where $R_{ij}$ correspond to the elementary rotations in the
$ij=23$, $13$, and $12$ planes (parametrized by three mixing angles
$c^{}_{ij} \equiv \cos \theta^{}_{ij}$ and $s^{}_{ij} \equiv \sin
\theta^{}_{ij}$), $P_{\delta}={\rm diag}(1,1,{\rm e}^{{\rm
i}\delta})$, and $P_{M}={\rm diag}({\rm e}^{{\rm i}\phi_1}, {\rm
e}^{{\rm i}\phi_2},1)$ contain the Dirac and Majorana CP-violating
phases, respectively.

According to Eq.~\eqref{eq:m}, the Yukawa coupling matrix $h$ is
related to the light neutrino mass matrix as $h=m/(\sqrt{2}v_{\rm
L})$. Hence, the flavor structure of the Yukawa coupling $h$ is
identical to that of the neutrino mass matrix, which allows for a
direct test of neutrino parameters via measurements of $h$. For
example, if kinematically accessible, the doubly charged component
of the Higgs triplet can be on-shell produced at colliders, and its
subsequent LNV or LFV decays may bring in significant signatures
allowing the determination of flavor structure of $h$, and therefore
$m$, at current and forthcoming colliders. In particular, the decay
branching ratios of $\Delta$ are shown to be highly sensitive to the
neutrino mixing parameters and the neutrino mass hierarchy.

The possibility of testing the neutrino mass matrix by studying the
decays of Higgs triplet at hadron colliders, e.g., the LHC, has been
discussed
intensively~\cite{Chun:2003ej,Akeroyd:2005gt,Garayoa:2007fw,Kadastik:2007yd,Akeroyd:2007zv,Chen:2008jh,Ren:2008yi,Petcov:2009zr,Nishiura:2009jn}.
In what follows, we will concentrate on the production of the doubly
charged Higgs $\Delta^{--}$ at linear $e^-e^-$ colliders, and in
particular, investigate the LFV processes $e^- e^- \rightarrow
\alpha^- \beta^-$ in detail.

\section{Doubly charged Higgs at a linear $e^-e^-$ collider}
\label{sec:collider signatures}

\begin{figure}[t]
\begin{center}\vspace{0.1cm}
\includegraphics[width=8cm]{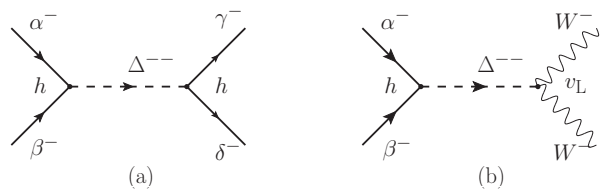}
\vspace{0cm} \caption{\label{fig:fig1} Feynman diagrams for the
doubly charged Higgs mediated bilepton channel (left) and $WW$
channel (right).}
\end{center}\vspace{-0.3cm}
\end{figure}

Among various interesting physics possibilities at a lepton-lepton
collider, one of the most promising processes to be investigated is
the bilepton production mediated by a doubly charged scalar
$\Delta^{--}$. The production of Higgs triplets in like-sign lepton
collisions has been discussed in
Refs.~\cite{Rizzo:1981xx,Rizzo:1982kn,London:1987nz,Frampton:1991mt,Barger:1994wa,Gluza:1995ix,Gluza:1995ky,Belanger:1995nh,Gunion:1995mq,Heusch:1996an,Gluza:1997kg,Cuypers:1997qg,Raidal:1997tb,Duka:1998fv,Mukhopadhyaya:2005vf,Rodejohann:2010jh}.
The relevant diagram is shown in the left-hand panel of
Fig.~\ref{fig:fig1}.
%%%%%%%%%%%%%%%%%%%%%%%%%%%%%%%%%%%%%%%%%%%
%%%%%%%%%%%%%%%%%%%% Fig.~1 %%%%%%%%%%%%%%%
%%%%%%%%%%%%%%%%%%%%%%%%%%%%%%%%%%%%%%%%%%%
In the framework under discussion, the
general cross section for $\alpha^- \beta^- \rightarrow \gamma^- \delta^-$,
where $\alpha, \beta, \gamma, \delta = e, \mu, \tau$,
reads
\begin{eqnarray}
\sigma & =& \frac{|h_{\alpha\beta}
h_{\gamma\delta}|^2}{4\pi(1+\delta_{\gamma\delta})}
\frac{s}{(s-m^2_\Delta)^2+m^2_\Delta \Gamma^2_\Delta} \nonumber
\\ &=&
\frac{|h_{\alpha\beta}h_{\gamma\delta}|^2}{(1+\delta_{\gamma\delta})}
\sigma_0 \, , \label{eq:crossll}
\end{eqnarray}
where $\Gamma_\Delta$ is the decay width of $\Delta^{--}$, and the
bare cross section $\sigma_0$ is independent of the
flavor indices. Note that, if there are two electrons in the final
states, the above formula does not apply, since the SM contributions
mediated by $\gamma$ and $Z$ may play the dominating role.

In addition to the bilepton channel, the inverse neutrinoless double
beta decay process $e^-e^- \rightarrow W^-W^-$ can also be used as a
probe of
LNV~\cite{Rizzo:1981xx,Rizzo:1982kn,London:1987nz,Frampton:1991mt,Barger:1994wa,Gluza:1995ix,Gluza:1995ky,Belanger:1995nh,Gunion:1995mq,Heusch:1996an,Gluza:1997kg,Cuypers:1997qg,Raidal:1997tb,Duka:1998fv,Mukhopadhyaya:2005vf,Rodejohann:2010jh}.
The corresponding diagram is shown in the right-hand panel of
Fig.~\ref{fig:fig1}. The cross section is given by
\begin{eqnarray}
\sigma & = & \nonumber \frac{G^2_F v^2_{\rm L}}{2\pi}
\left|h_{\alpha\beta}\right|^2 \frac{(s-2m^2_W)^2 +
8m^4_W}{(s-m^2_\Delta)^2 + m^2_\Delta \Gamma^2_\Delta} \sqrt{1-4
\frac{m^2_W}{s}}
\\ & \simeq & \frac{G^2_F v^2_{\rm L}}{2\pi}
\left|h_{\alpha\beta}\right|^2
\frac{s^2}{(s-m^2_\Delta)^2+m^2_\Delta \Gamma^2_\Delta} \, ,
\label{eq:crossWW}
\end{eqnarray}
where the mass of $W$-boson $m_W$ is neglected in the second line.
Depending on the mass splitting within the Higgs triplet, the decays
$\Delta^{--} \rightarrow W^- \Delta^-$ and $\Delta^{--} \rightarrow
\Delta^-\Delta^-$ may also take place, which may be dominating as
they are driven by a gauge coupling. Here we assume a degeneracy
among the Higgs triplet components, viz., the above two channels are
kinematically suppressed. Accordingly, the relative ratio of the
cross sections can be estimated by
\begin{eqnarray}
\frac{\sigma(\alpha^- \beta^- \rightarrow W^-W^-)}{\sigma(\alpha^-
\beta^- \rightarrow \gamma^- \delta^-)} & \simeq & \frac{2G^2_F
v^2_{\rm L}
(1+\delta_{\gamma\delta})s}{\left|h_{\gamma\delta}\right|^2}
\nonumber \\
 =  \frac{\Gamma(\Delta^{--}\rightarrow W^-W^-)}{\Gamma(\Delta^{--}\rightarrow
\gamma^- \delta^-)} \frac{s}{m_\Delta} \, . & &
\label{eq:crossratio}
\end{eqnarray}
The requirement $\sigma(\alpha^- \beta^- \rightarrow W^-W^-) \ll
\sigma(\alpha^- \beta^- \rightarrow \gamma^- \delta^-)$ implies that
the bilepton decays dominate the decay of $\Delta^{--}$, which
occurs when $v_{\rm L} < 10^{-4}~{\rm GeV}$ for a TeV scale Higgs
triplet~\cite{Perez:2008ha}.

Inserting Eq.~\eqref{eq:m} into Eq.~\eqref{eq:crossll}, one can
rewrite the cross section in terms of the neutrino mass matrix
elements as
\begin{eqnarray}
\sigma(\alpha^- \beta^- \rightarrow \gamma^- \delta^-) =
\frac{\left|m_{\alpha\beta}m_{\gamma\delta}\right|^2}{4v^4_{\rm
L}(1+\delta_{\gamma\delta})}\sigma_0\, . \label{eq:cross2}
\end{eqnarray}
One can then observe that the cross sections are correlated to both
the neutrino mass matrix and $v_{\rm L}$. If the neutrino mass scale
is fixed, a smaller $v_{\rm L}$ generally corresponds to larger
Yukawa couplings and hence larger cross sections. In case of the
$e^-e^-$ collider, the above equation reduces to
\begin{eqnarray}
\sigma(e^- e^- \rightarrow \gamma^- \delta^-) =
\frac{\left|m_{ee}\right|^2\left|m_{\gamma\delta}\right|^2}{4v^4_{\rm
L}(1+\delta_{\gamma\delta})}\sigma_0\, , \label{eq:crossee}
\end{eqnarray}
where $\left|m_{ee}\right|$ is the effective mass of the
neutrinoless double beta decay process\footnote{Note that
neutrinoless double beta decay is also triggered by $\Delta^{--}$
via the reverse diagram of Fig.~\ref{fig:fig1}b. However, the
contribution from the Higgs triplet is suppressed by a factor
$q^2/m^2_\Delta$ ($q$ represents the momentum transfer carried by
the exchanged neutrinos, and is typically of order $10~{\rm MeV}$)
compared to the standard contribution from a Majorana mass term of
light neutrinos.}, and constrained by current experiments as
$\left|m_{ee}\right| \lesssim 1~{\rm eV}$. If the effective mass of
neutrinoless double beta decay turns out to be very small, there
will be no visible collider signatures either. Then the process
might be observed in running a future muon collider in the $\mu^-
\mu^-$ mode, or in an $e^- \mu^-$ collider.

%%%%%%%%%%%%%%%%%%%%%%%%%%%%%%%%%%%%%%%%%%%%%%%%
%%%%%%%%%%%%%%%%%%%%%%%%%%%%%%%%%%%%%%%%%%%%%%%%
\begin{table}[t]
\begin{center}
\vspace{0.1cm}
\begin{tabular}{|c|c|c|}
%-------------------------------------------------
\hline Decay & Constraint on & Bound (90~\%~C.L.)
\\
\hline
$\mu^- \rightarrow e^- e^+ e^- $  &  $|h_{ee}h_{e\mu}|^2 \left(\frac{250~{\rm GeV}}{m_\Delta}\right)^4$ & $2.1 \times 10^{-12}$  \\
\hline $\tau^-
\rightarrow e^- e^+  e^- $  & $|h_{ee}h_{e\tau}|^2 \left(\frac{250~{\rm GeV}}{m_\Delta}\right)^4$   & $ 4.4 \times 10^{-7}$  \\
\hline
$\tau^- \rightarrow  \mu^- \mu^+ \mu^-$  & $|h_{\mu\mu}h_{\mu\tau}|^2 \left(\frac{250~{\rm GeV}}{m_\Delta}\right)^4$  & $ 3.9 \times 10^{-7}$  \\
\hline
$\tau^- \rightarrow e^- \mu^+  e^-$  & $|h_{ee}h_{\mu\tau}|^2 \left(\frac{250~{\rm GeV}}{m_\Delta}\right)^4$   & $ 2.4 \times 10^{-7}$  \\
\hline
$\tau^- \rightarrow \mu^- e^+  \mu^-$  & $|h_{e\mu}h_{e\tau}|^2 \left(\frac{250~{\rm GeV}}{m_\Delta}\right)^4$   &  $ 2.8 \times 10^{-7}$  \\
\hline
$\tau^- \rightarrow e^- \mu^+ \mu^-  $  & $|h_{e\mu}h_{\mu\tau}|^2 \left(\frac{250~{\rm GeV}}{m_\Delta}\right)^4$  &  $ 2.3 \times 10^{-7}$ \\
\hline
$\tau^- \rightarrow e^- e^+ \mu^-$  & $|h_{\mu\mu}h_{e\tau}|^2 \left(\frac{250~{\rm GeV}}{m_\Delta}\right)^4$ &  $ 1.7 \times 10^{-7}$ \\
\hline
$\mu^- \rightarrow e^- \gamma  $  &  $|(hh^\dagger)_{e\mu}|^2 \left(\frac{250~{\rm GeV}}{m_\Delta}\right)^4$ &  $ 6.5 \times 10^{-9}$ \\
\hline
$\tau^- \rightarrow e^- \gamma  $  & $|(hh^\dagger)_{e\mu}|^2 \left(\frac{250~{\rm GeV}}{m_\Delta}\right)^4$ &  $ 1.0 \times 10^{-4}$ \\
\hline
$\tau^- \rightarrow \mu^- \gamma  $  & $|(hh^\dagger)_{e\mu}|^2 \left(\frac{250~{\rm GeV}}{m_\Delta}\right)^4$ &  $ 1.4 \times 10^{-4}$ \\
%-------------------------------------------------
\hline
$\mu^+ e^- \rightarrow \mu^- e^+$  & $|h_{ee}h_{\mu\mu}|^2 \left(\frac{250~{\rm GeV}}{m_\Delta}\right)^4$  &  $ 9.5 \times 10^{-6}$ \\
%-------------------------------------------------
\hline
%-------------------------------------------------
\end{tabular}
\end{center}
\caption{Constraints (at 90~\%~C.L.) on $h$ from $\ell \rightarrow
\ell\ell\ell$, $\ell \rightarrow \ell \gamma$, and $\mu^+
e^- \rightarrow \mu^- e^+$ processes. The experimental bounds have
been obtained from Refs.~\cite{Willmann:1998gd,Amsler:2008zz}.}
\vspace{0.cm} \label{tab:constraints}
\end{table}
%%%%%%%%%%%%%%%%%%%%%%%%%%%%%%%%%%%%%%
%%%%%%%%%%%%%%%%%%%%%%%%%%%%%%%%%%%%%%

The most relevant experimental constraints on the Yukawa coupling
matrix $h$ come from the LFV decays $\mu\rightarrow 3e$ and
$\tau\rightarrow 3\ell$ (which occur a tree level), the radiative
lepton decays $\ell_\alpha \rightarrow \ell_\beta \gamma$
(one-loop), and the muonium to antimuonium conversion. Bounds from
Bhabha scattering and universality tests of weak interactions are
relatively weaker and will be not elaborated on in our calculations.
The constraints at 90~\%~C.L.~are summarized in
Table~\ref{tab:constraints} (see also
Refs.~\cite{Cuypers:1996ia,Frampton:1997gc,Akeroyd:2009nu}).

%%%%%%%%%%%%%%%%%%%%%%%%%%%%%%%%%%%%%%%%%%%
%%%%%%%%%%%%%%%%%%%% Fig.~2 %%%%%%%%%%%%%%%
\begin{figure*}[t]
\begin{center}\vspace{-0.2cm}
\includegraphics[width=5.8cm]{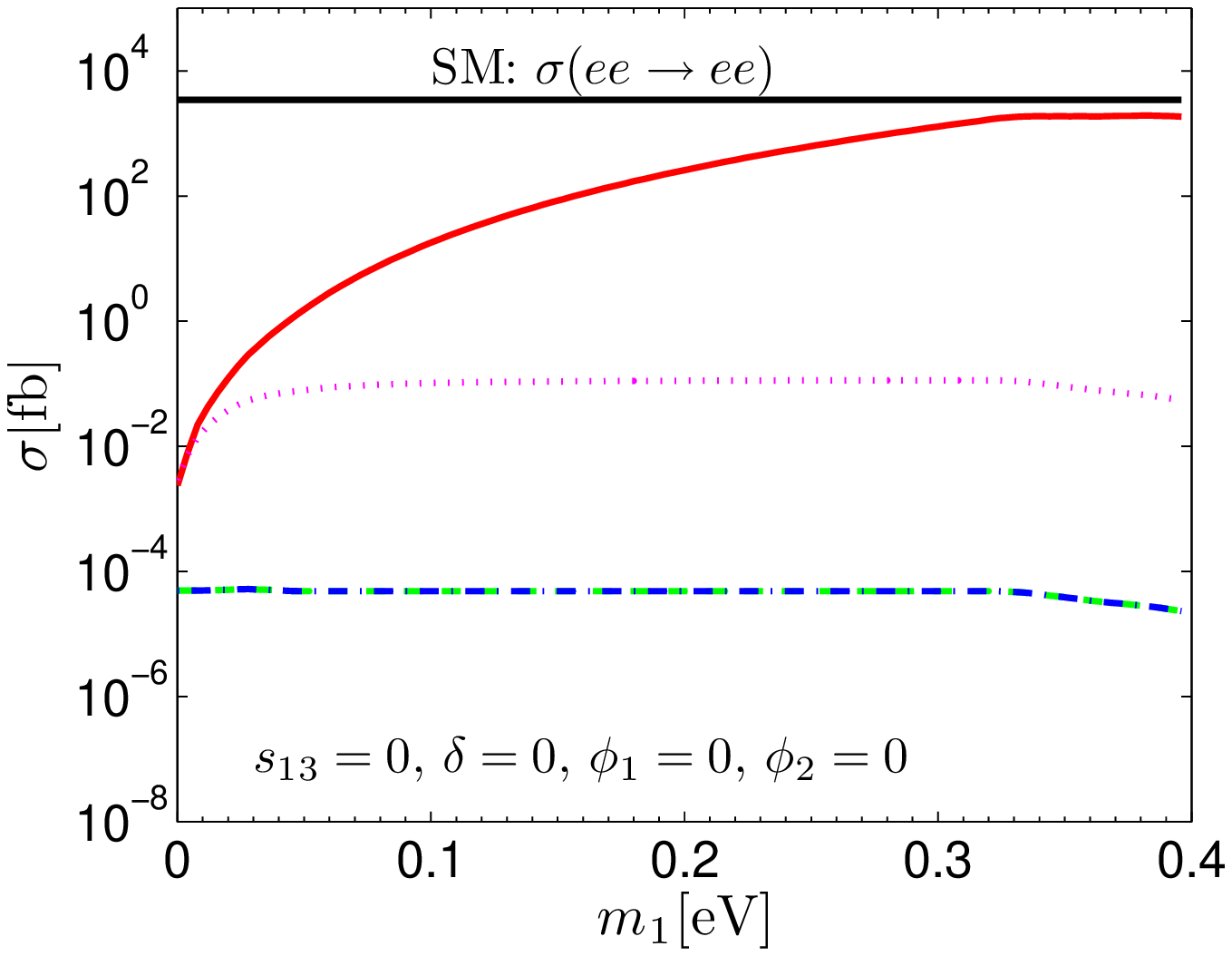}
\includegraphics[width=5.8cm]{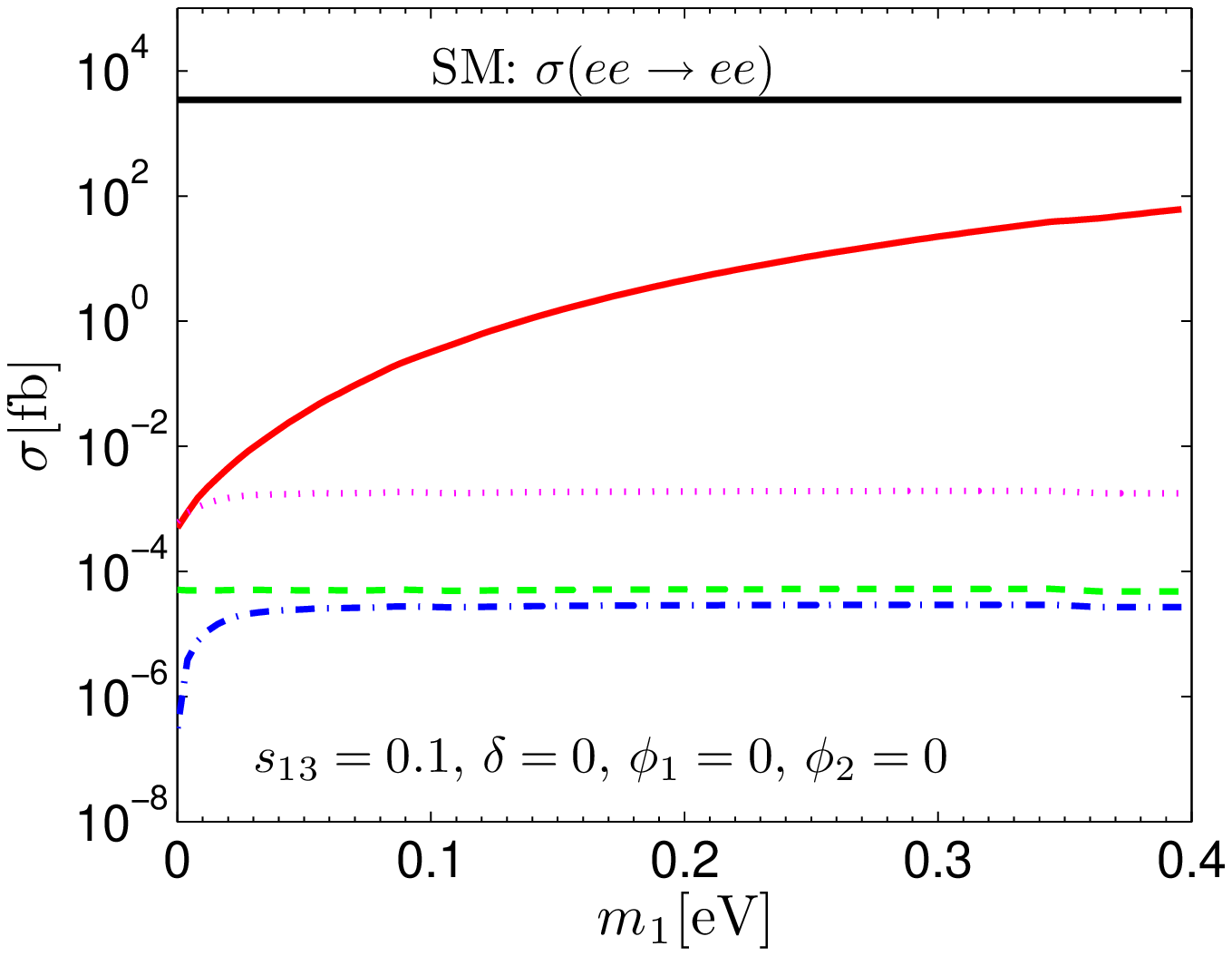}
\includegraphics[width=5.8cm]{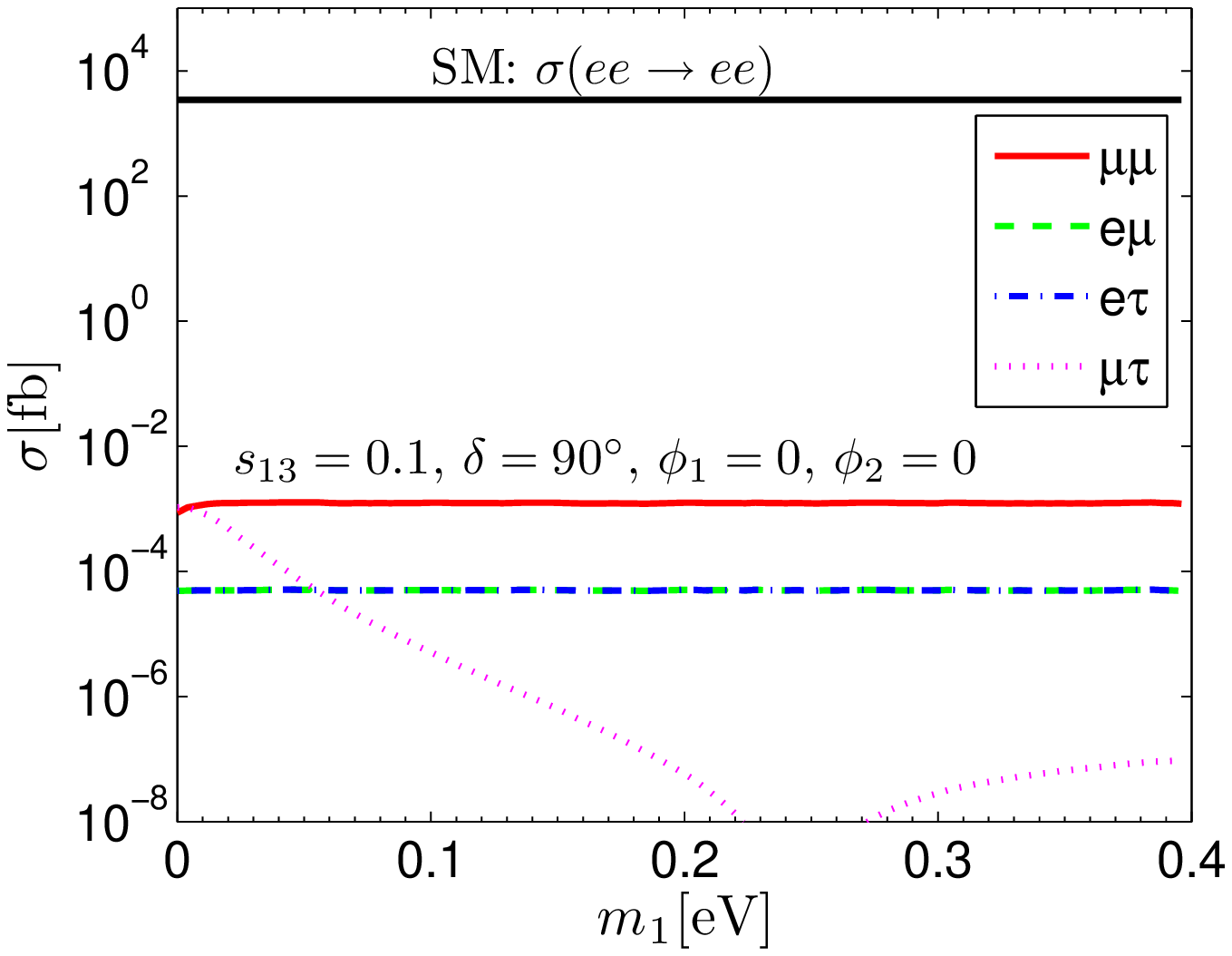}
\includegraphics[width=5.8cm]{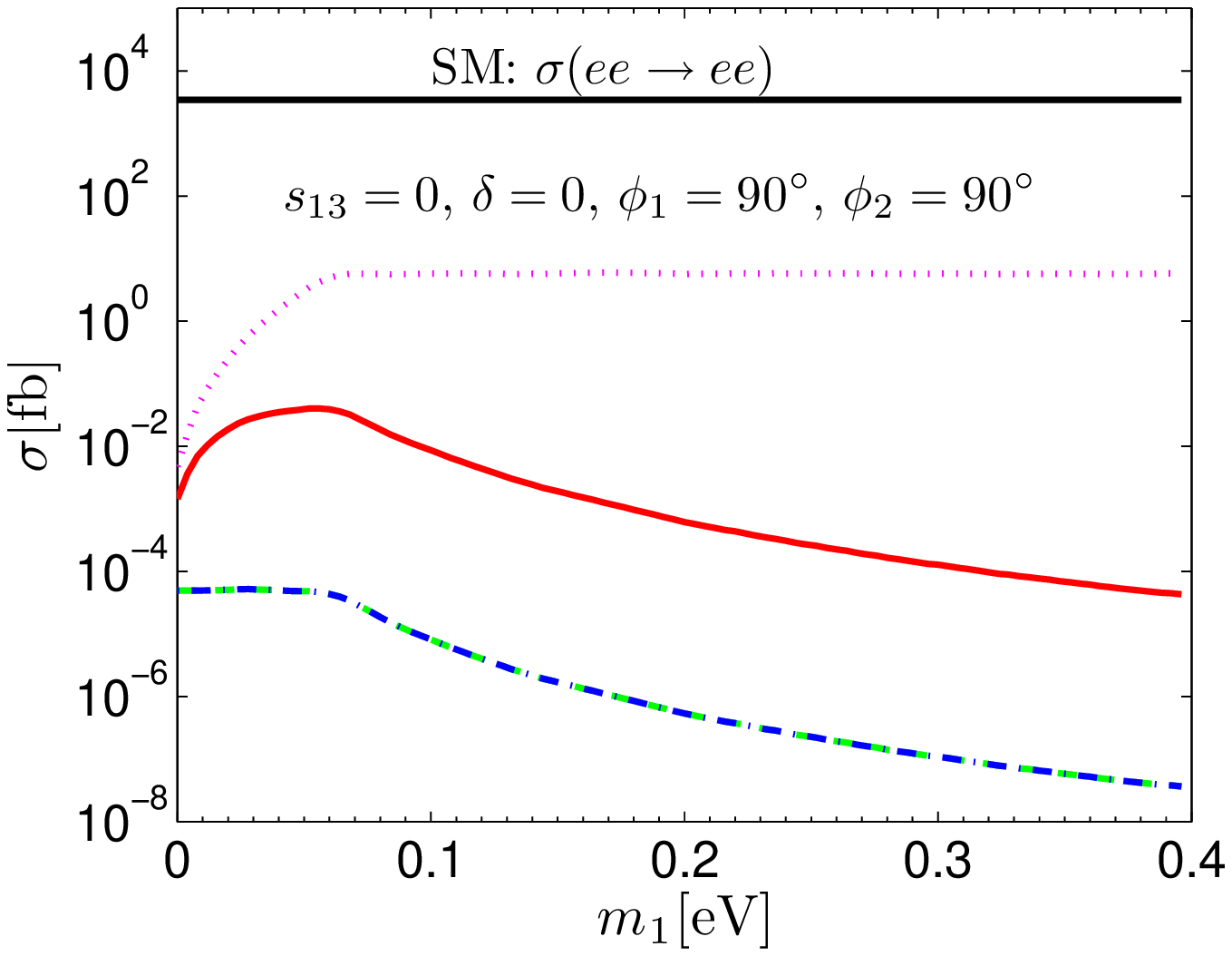}
\includegraphics[width=5.8cm]{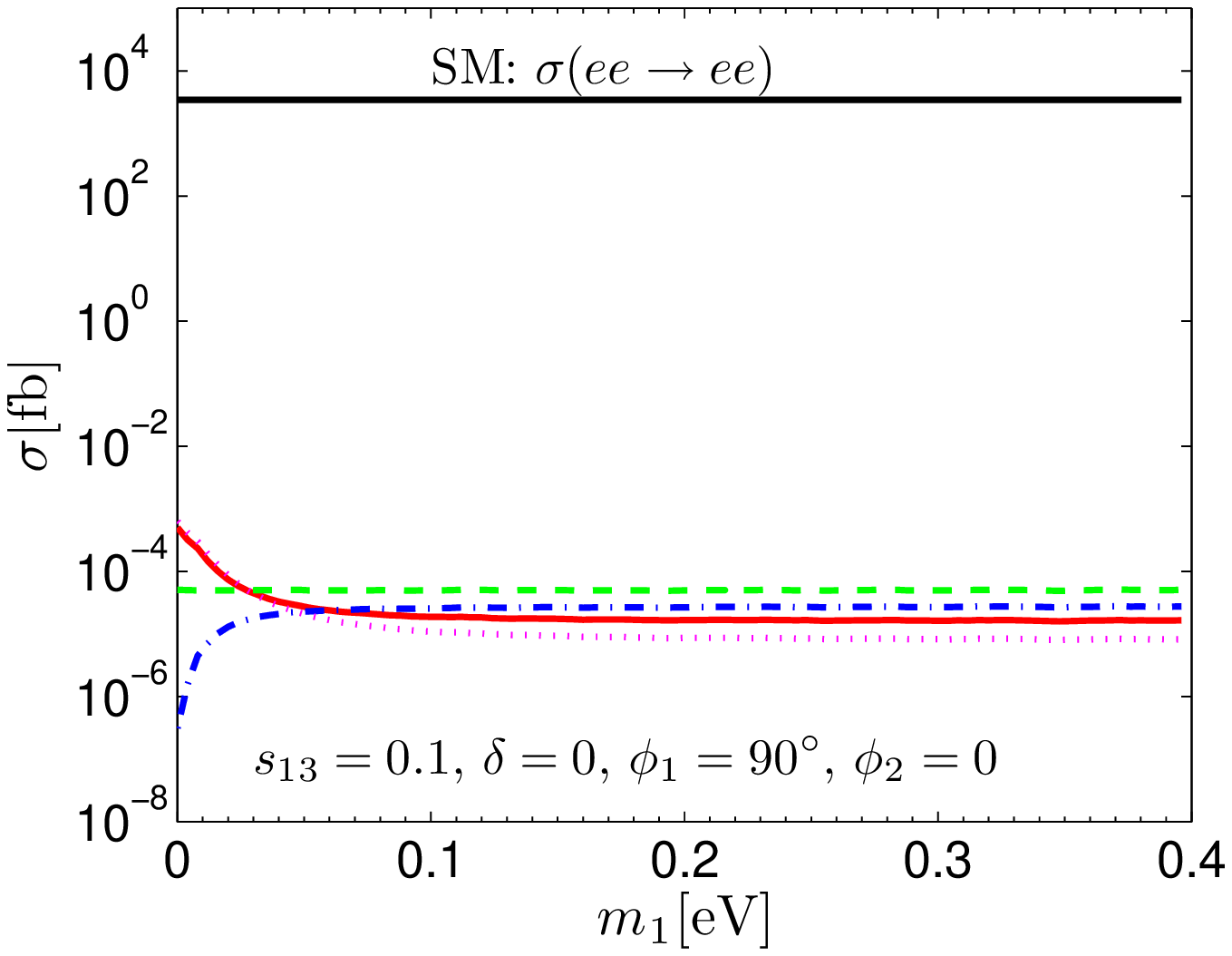}
\includegraphics[width=5.8cm]{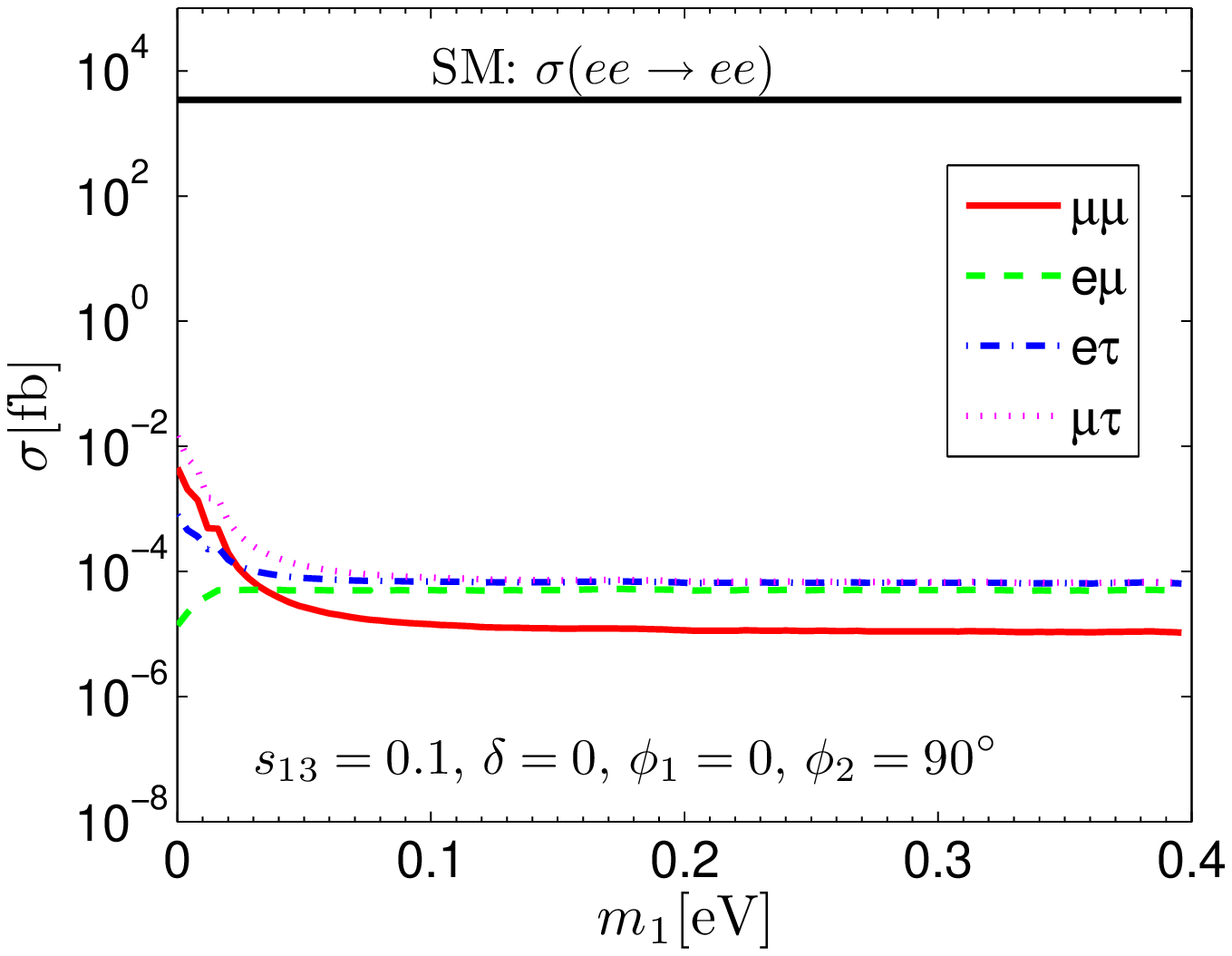}
\end{center}
\vspace{-0.3cm} \caption{\label{fig:cross} Upper limits of cross
sections $e^-e^- \rightarrow \alpha^- \beta^-$ with respect to $m_1$
at a linear collider with $\sqrt{s}=1~{\rm TeV}$. The Higgs triplet
mass $m_\Delta = 800~{\rm GeV}$ has been assumed as an example,
while the choices of $\sin\theta_{13}$ are labeled on each plot.
Furthermore, we take the Majorana CP-violating phases $\phi_1$ and
$\phi_2$ to be zero for the plots on the upper panel, and the Dirac
CP-violating phase $\delta=0$ for the plots on the lower panel. For
comparison, the SM M{\o}ller scattering cross section is also shown
on the plots with black solid lines. The luminosity one can assume
is 80 fb$^{-1}$.} \vspace{0.cm}
\end{figure*}
%%%%%%%%%%%%%%%%%%%%%%%%%%%%%%%%%%%%%%%%%%%
%%%%%%%%%%%%%%%%%%%%%%%%%%%%%%%%%%%%%%%%%%%

%%%%%%%%%%%%%%%%%%%%%%%%%%%%%%%%%%%%%%%%%%%
%%%%%%%%%%%%%%%%%%%% Fig.~3 %%%%%%%%%%%%%%%
\begin{figure*}[p]
\begin{center}\vspace{-0.8cm}
\includegraphics[width=5.7cm,bb=0 0 700 700]{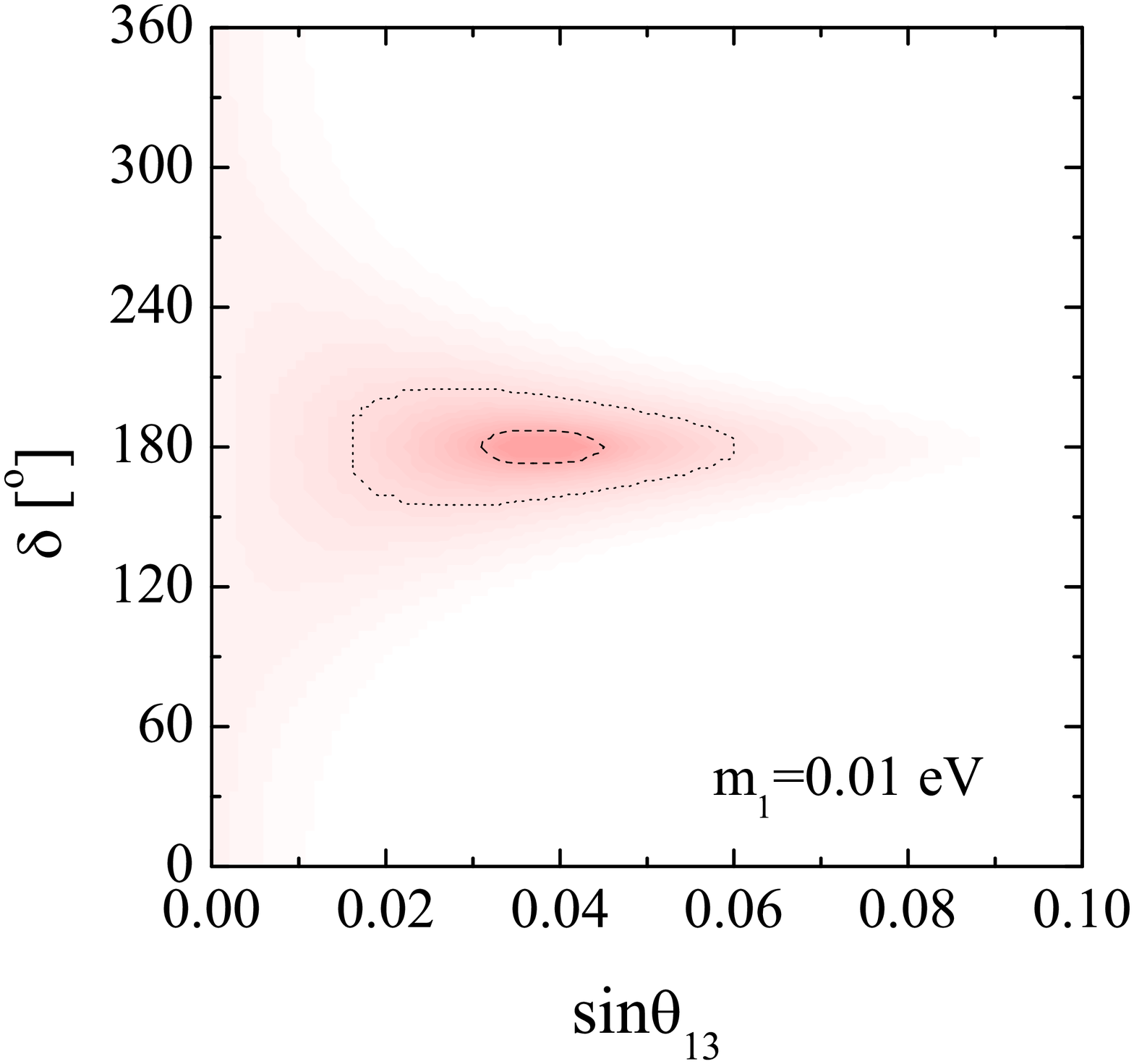}
\includegraphics[width=5.7cm,bb=50 0 750 700]{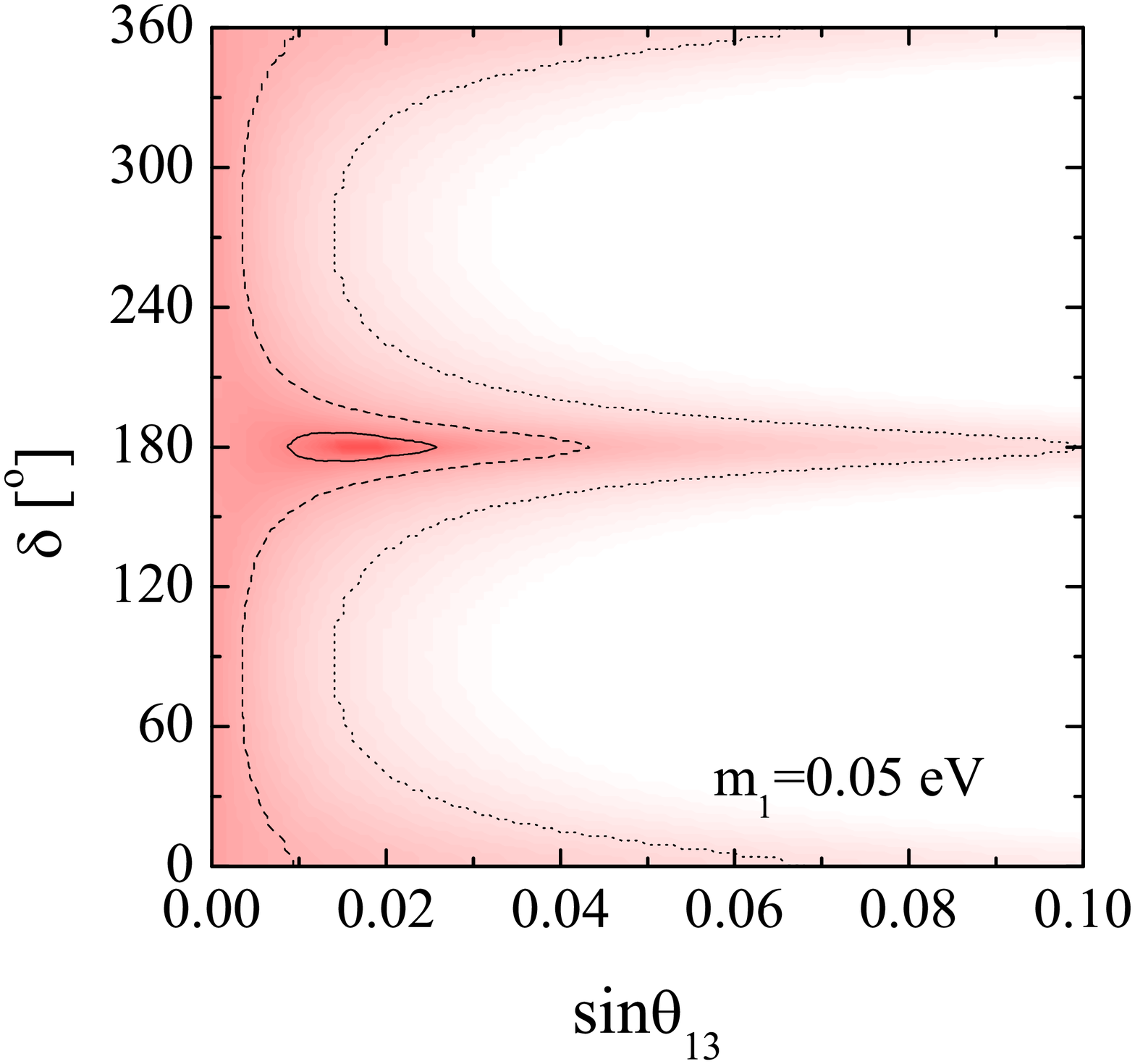}
\includegraphics[width=5.7cm,bb=100 0 800 700]{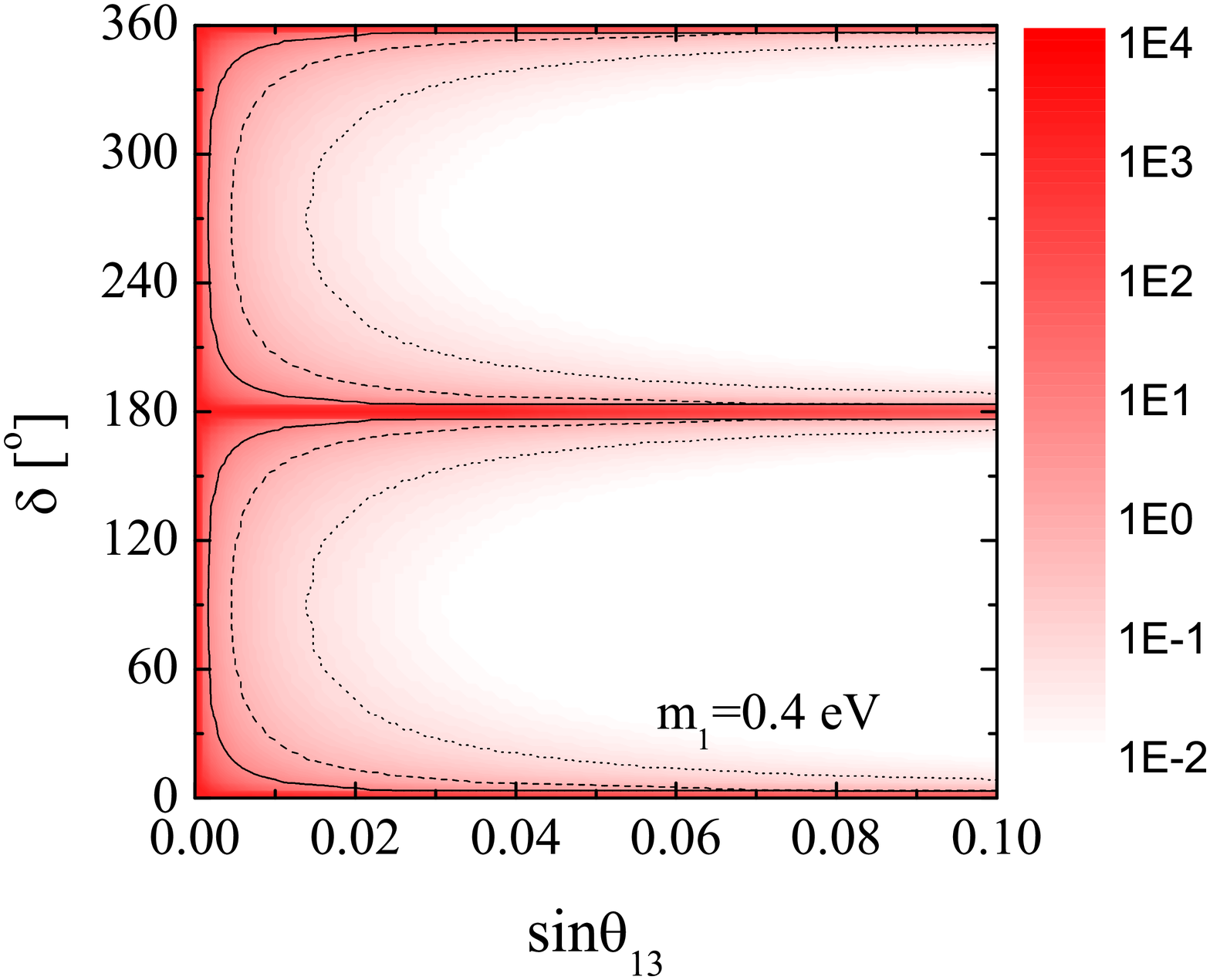} \\
\vspace{-0.3cm}
\includegraphics[width=5.7cm,bb=0 -80 700 620]{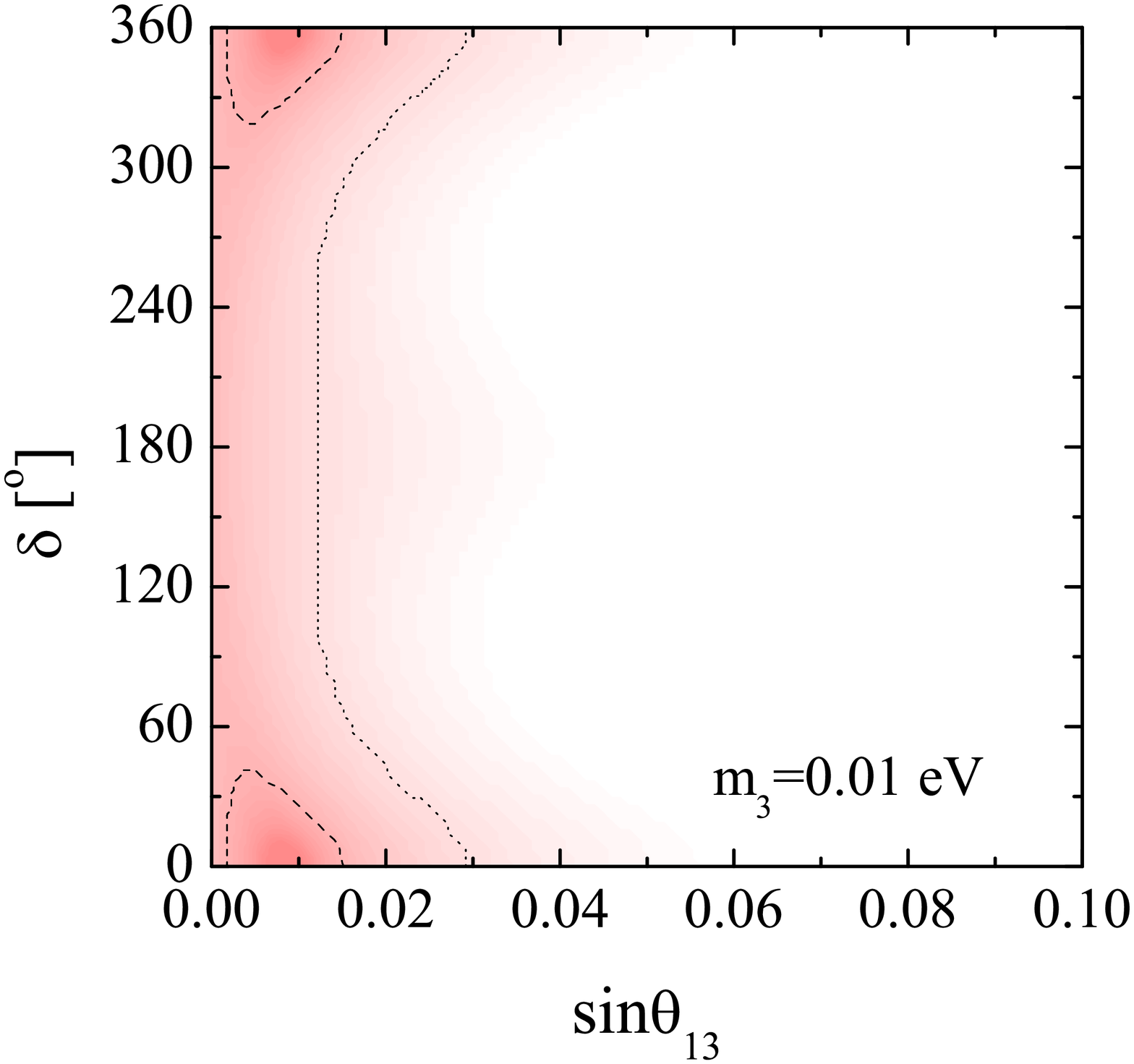}
\includegraphics[width=5.7cm,bb=50 -80 750 620]{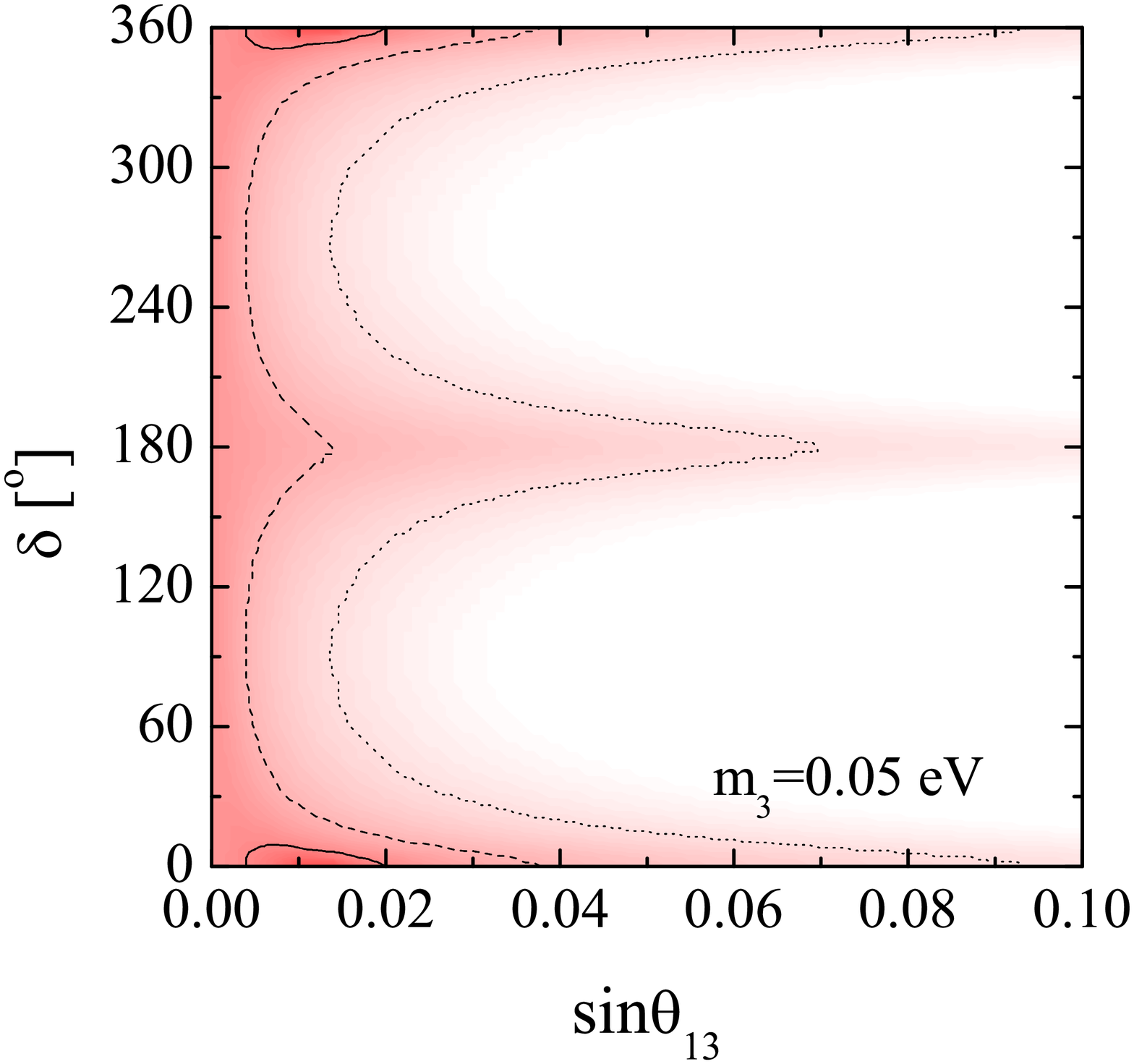}
\includegraphics[width=5.7cm,bb=100 -80 800 620]{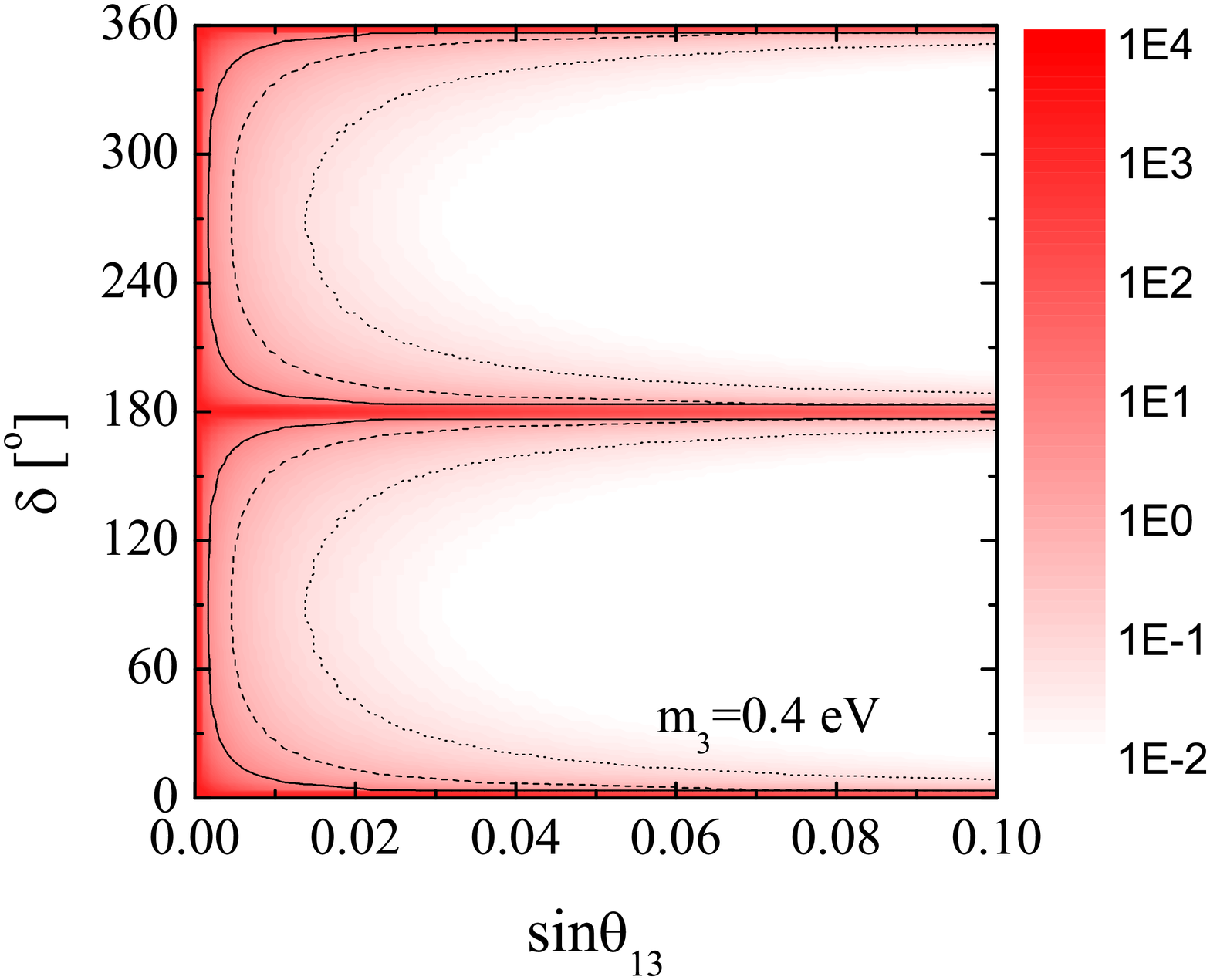}
\vspace{-0.7cm}\caption{\label{fig:contour} Upper bounds of the
total cross section $\sigma(e^-e^-\rightarrow \mu^-\mu^-)$ (in units
of fb) for $\sqrt{s} = 1$ TeV in the $\sin\theta_{13}-\delta$ plane
with the experimental setup and the mass of Higgs triplet are the
same as these in Fig.~\ref{fig:cross}. For the plots in the upper
row, we consider the normal mass ordering, while in the lower row
the inverted mass ordering is assumed. The lightest neutrino mass
can be read off from the plots, and for the sake of simplicity, we
take all the Majorana phases to be zero. For an integrated
luminosity $80~{\rm fb}^{-1}$, the solid, dashed and dotted curves
show the regions with expected number of events greater than 5, 50
and 500, respectively.}
%%%%%%%%%%%%%%%%%%%%%%%%%%%%%%%%%%%%%%%%%
%%%%%%%%%%%%%%%%%%%%% Fig.~4 %%%%%%%%%%%%%%%
\vspace{-0.1cm}
\includegraphics[width=5.7cm,bb=0 0 700 700]{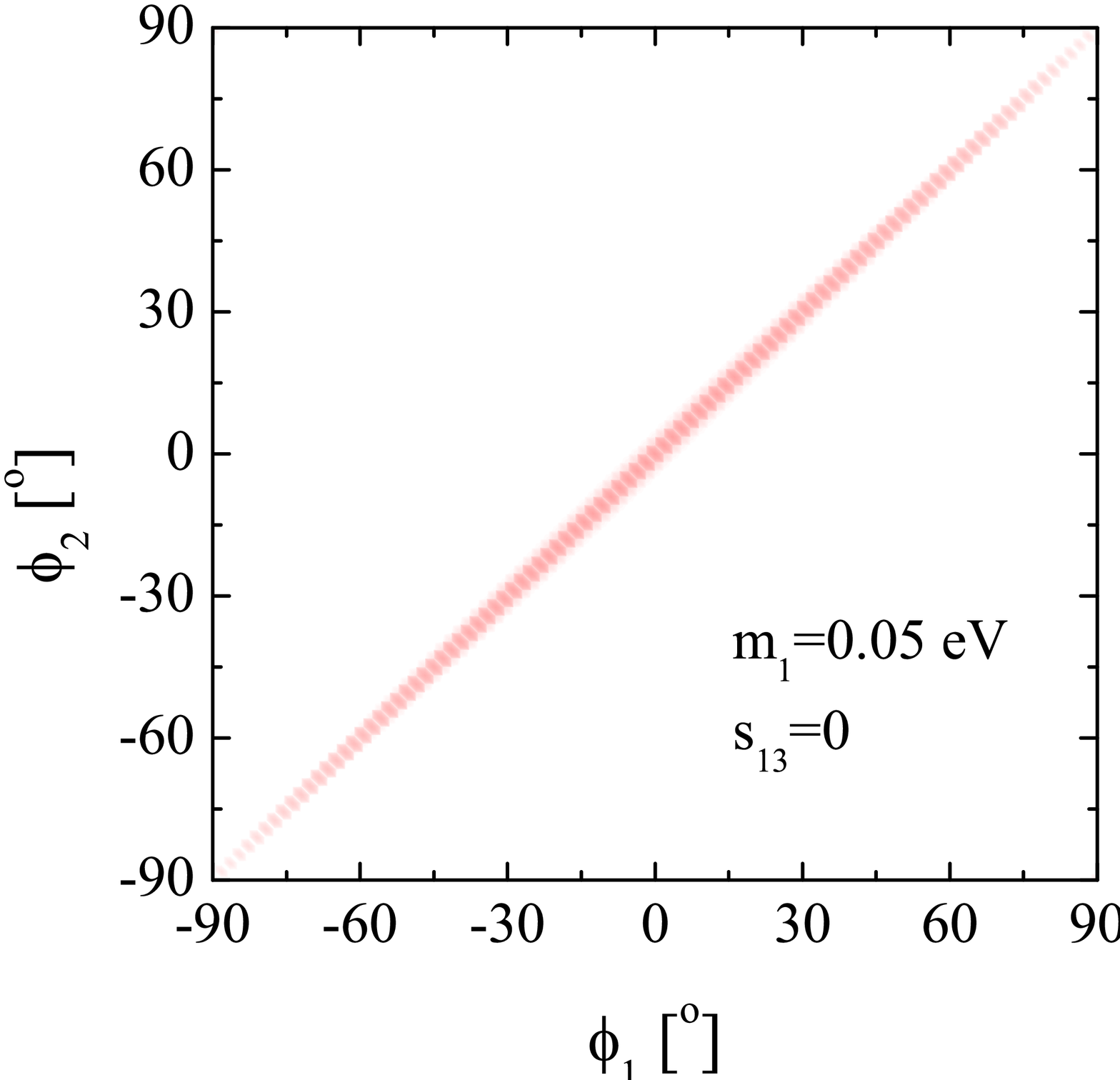}
\includegraphics[width=5.7cm,bb=50 0 750 700]{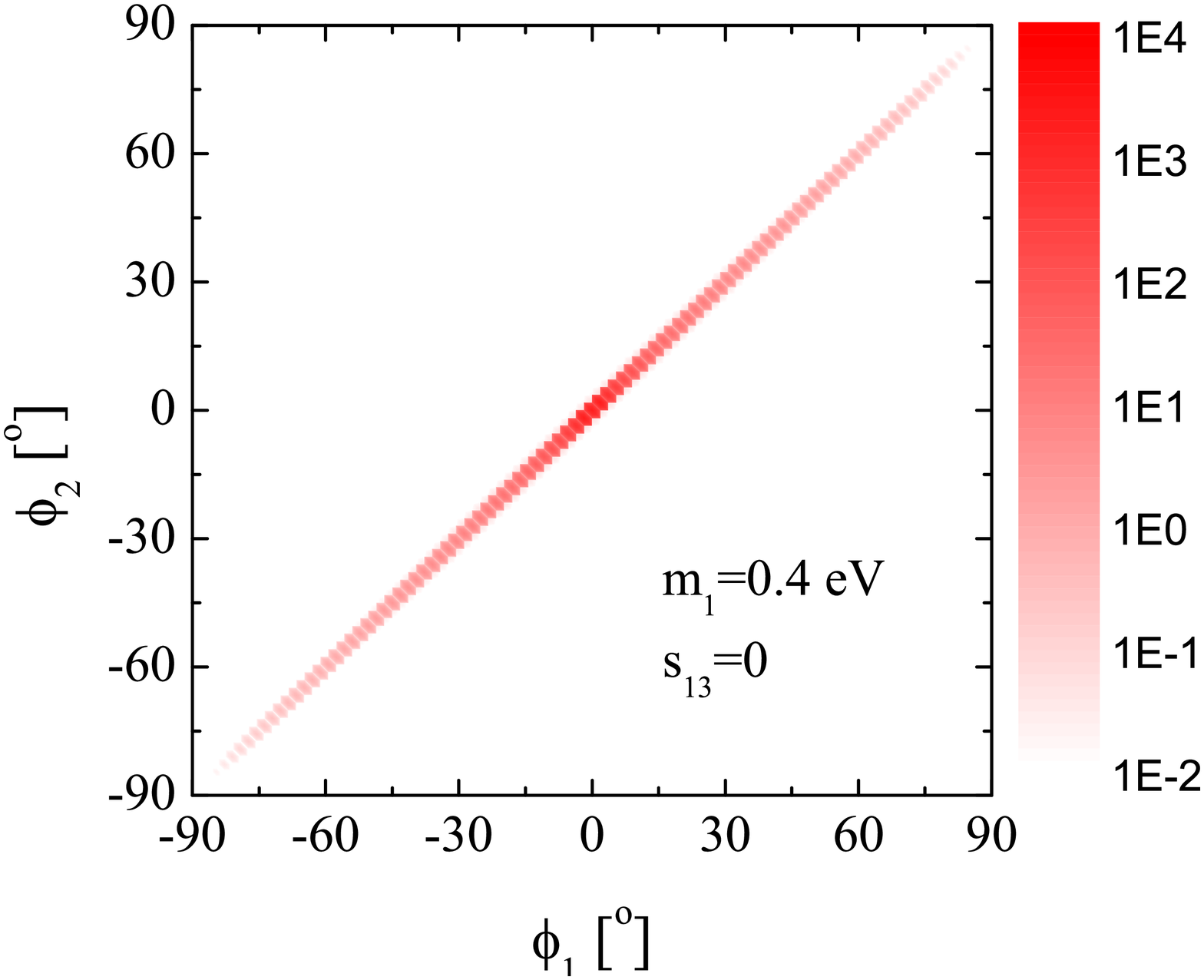}\\
\vspace{-0.8cm}
\includegraphics[width=5.7cm,bb=0 0 700 700]{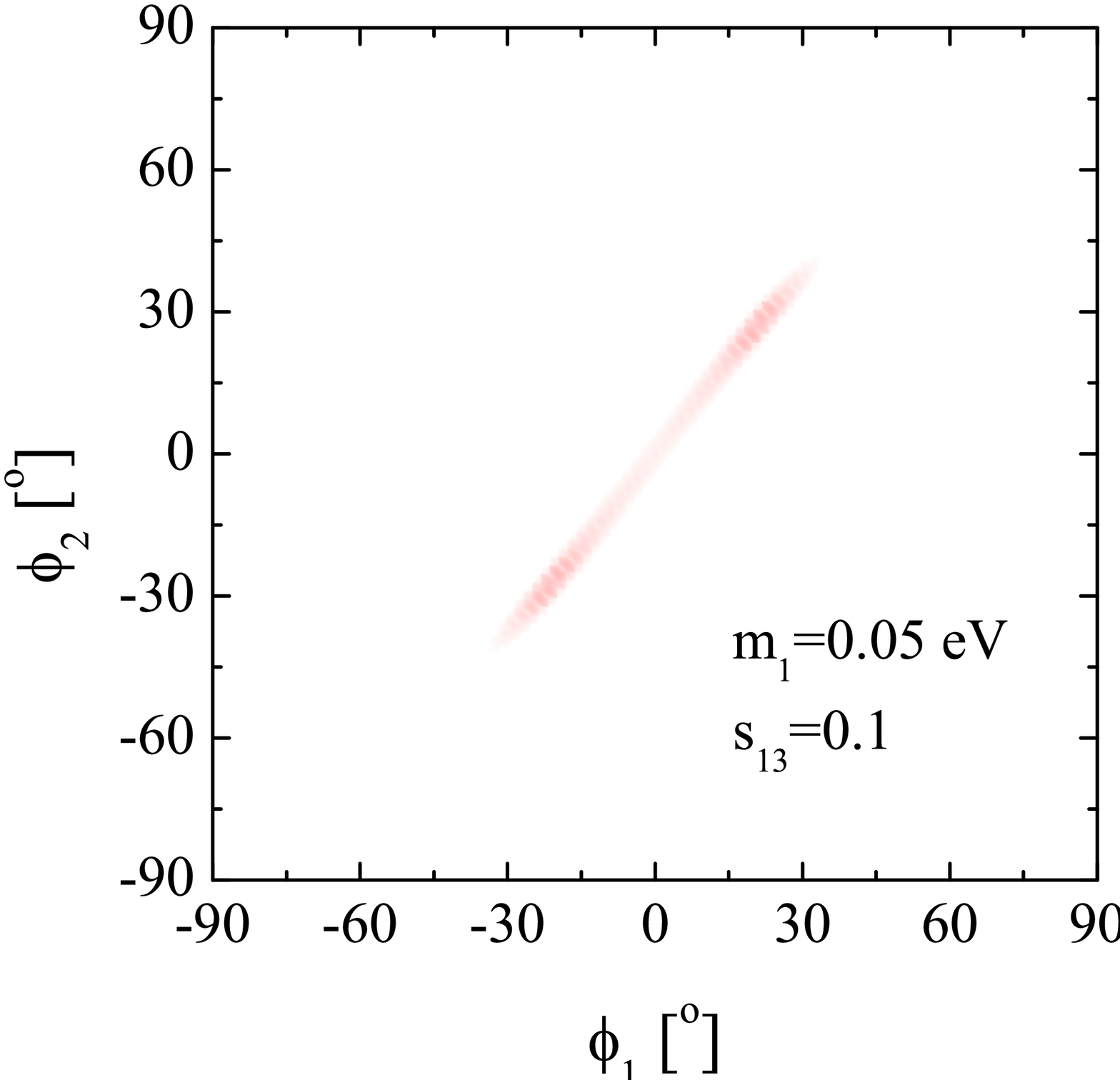}
\includegraphics[width=5.7cm,bb=50 0 750 700]{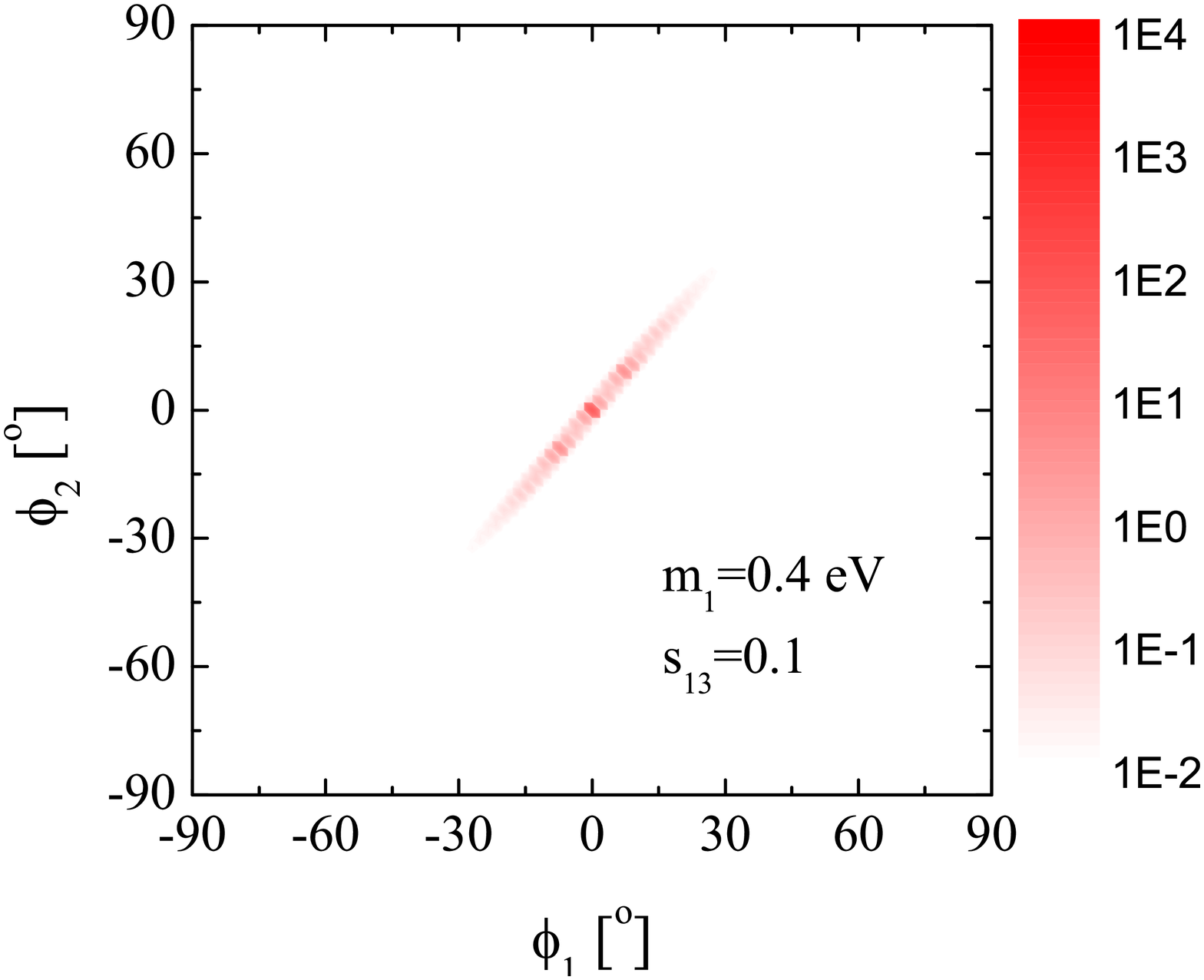}
\vspace{0.2cm} \caption{\label{fig:phase} Upper bounds of the total
cross section $\sigma(e^-e^-\rightarrow \mu^-\mu^-)$ (in units of
fb) in the $\phi_1-\phi_2$ plane with the same experimental setting
of Fig.~\ref{fig:contour}. The input values of $s_{13}$ and $m_1$
are labeled on the plots, while $\delta$ is taken to be zero. The
luminosity one can assume is 80 fb$^{-1}$.
 }
\end{center}\vspace{0cm}
\end{figure*}
%%%%%%%%%%%%%%%%%%%%%%%%%%%%%%%%%%%%%%%%%
%%%%%%%%%%%%%%%%%%%%%%%%%%%%%%%%%%%%%%%%%

An interesting feature is that, among all the relevant experimental
bounds, at least one off-diagonal element in $h$ is involved except
for the constraint from muonium-antimuonium conversion. Note further
that the most stringent experimental constraint comes from the rare
muon decays $\mu\rightarrow 3e$ and $\mu \rightarrow e\gamma$. In
order to generate observable collider signatures and avoid large LFV
processes at the same time, one requires the flavor non-diagonal
parts of $h$ to be relatively small, in particular, the $e\mu$
component\footnote{Single texture zeros in the neutrino mass matrix
have been studied in Ref.~\cite{Merle:2006du}.}. One way to achieve
this goal requires that $m$ takes an approximately diagonal form,
which is the case for a nearly degenerate (ND) neutrino mass
spectrum, i.e., $m_1 \simeq m_2 \simeq m_3 =m_0$, together with
vanishing CP-violating phases. For example, in the ND case and
neglecting the smallest mixing angle $\theta_{13}$, $h_{e\mu}$ is
given by
\begin{eqnarray}
\left|h_{e\mu}\right|^2 \simeq \frac{2m_0^2}{v^2_{\rm L}}
s^2_{12}c^2_{12} c^2_{23} \sin^2\left(\phi_1-\phi_2\right) \, .
\end{eqnarray}
Therefore, in the limit $\phi_1-\phi_2 \simeq 0$, the most stringent
constraint in Table~\ref{tab:constraints} disappears. This situation
holds no matter whether the neutrino mass ordering is normal
($m_1<m_2<m_3$) or inverted ($m_3<m_1<m_2$). Actually, to be precise
we should note that $h_{e\mu}$ cannot vanish exactly if
$\theta_{13}=0$ \cite{Merle:2006du}, but can be sufficiently small
for our purposes.

In the case of a hierarchical neutrino mass spectrum, i.e., $m_1
\simeq 0$ or $m_3 \simeq 0$, there still exists parameter space
ensuring the (close-to) vanishing of $h_{e\mu}$.
In the extreme case with $m_1
=0$, one can expand $h_{e\mu}$ according to small quantities
$s_{13}$ and $r=m_2/m_3$. Taking $\theta_{23}=\pi/4$, we obtain
approximately
\begin{eqnarray}
\left|h_{e\mu}\right|^2 \simeq \frac{m_2 m_3 s_{12}c_{12}}{4v^2_{\rm
L}} \left[r s_{12}c_{12} + 2 s_{13}c_{\delta+2\phi_2}\right] \, ,
\label{eq:normal-hem}
\end{eqnarray}
where higher order terms in proportion to $s^2_{13}$ or $r^2$ have
been neglected. Thus, the severe constraint from the $\mu\rightarrow
3e$ process is evaded if the relation
\begin{eqnarray}
r s_{12}c_{12} \simeq -2 s_{13}c_{\delta + 2\phi_2} \, ,
\label{eq:normal}
\end{eqnarray}
holds. As for the IH case, in the limit $m_3=0$, we obtain an
estimate
\begin{eqnarray}
\left|h_{e\mu}\right|^2 \simeq \frac{m^2_2
s^2_{12}c^2_{12}s^2_{\phi_1-\phi_2}}{v^2_{\rm L}} \, ,
\label{eq:inverted}
\end{eqnarray}
which again suggests degenerate Majorana CP-violating phases in
favor of observable collider signatures. We will see later that
typically the di-muon channel $e^- e^- \rightarrow \mu^- \mu^-$ is
of interest, and thus we want that both $h_{ee}$ and $h_{\mu\mu}$
are sizable. This implies again that nearly degenerate neutrinos,
and to some extent inversely hierarchical mass schemes will be
favored over a normal hierarchy, for which the $ee$ entry of the
mass matrix is typically much smaller than the $\mu\mu$ element.

\section{numerical results}
\label{sec:numerics}

In our numerical illustrations, we consider a linear $e^-e^-$
collider with center-of-mass energy $\sqrt{s}=1~{\rm TeV}$, while
the mass of Higgs triplet is assumed to be $m_\Delta = 800~{\rm
GeV}$. The typical luminosity is ${\cal L}=80~(\sqrt{s}/{\rm
TeV})^2~{\rm fb}^{-1}$. We also make use of the neutrino parameters
from a global-fit of the current neutrino oscillation
experiments~\cite{Schwetz:2008er}. In the numerical calculations, we
fix the neutrino parameters and $m_\Delta$. We let $v_{\rm L}$ vary
between $10^{-10} ~{\rm GeV}$ and $10^{-6}~{\rm GeV}$ (in order to
have the bilepton decays dominate over the $W^- W^-$ mode) and vary
the Yukawa coupling matrix $h$ until one of the upper bounds in
Table~\ref{tab:constraints} is saturated. Perturbativity of $h$ is
satisfied, and the Yukawas do not exceed 0.35 in all of the results
we will present in what follows.

In Fig.~\ref{fig:cross} we illustrate the upper limits of the cross
sections $\sigma(e^-e^- \rightarrow \alpha^- \beta^-)$ with respect
to the lightest neutrino mass $m_1$ given the constraints from
Table~\ref{tab:constraints}. The choices for the neutrino mixing
parameters are labeled on each plot, and the cross section of the SM
process $e^-e^-\rightarrow e^-e^-$ is also shown for the purpose of
comparison\footnote{In computing the total cross section in the SM
framework, $|\cos\theta|<0.8$ is used with $\theta$ being the
scattering angle between the initial and final electron in the
center-of-mass frame.}. Due to the experimental difficulty in tau
reconstruction, we do not show channels with two tau. As we
discussed in the previous section, the upper bounds of the cross
sections are very sensitive to the neutrino parameters. Especially,
in the ND case and vanishing CP-violating phases, e.g., the left
upper plot, the cross section of the channel with two muons in the
final states could be more than $1000~{\rm fb}$. It turns out that
for $m_1 \gtrsim 0.3~{\rm eV}$ the bound coming from $\mu^+e^-
\rightarrow \mu^-e^+$ is the dominating one\footnote{The next
generation muon factory may improve this constraint by one order of
magnitude~\cite{Nakahara:2009zz}, which however does not seriously
change the main conclusion addressed in this work.}, which then sets
an upper limit on the di-muon channel. Moreover, a non-zero $s_{13}$
will slightly decrease the maximal cross sections. One also observes
from plots in the right column that non-vanishing CP-violating
phases can suppress the di-muon cross section, mostly because
$h_{ee}$ and $h_{\mu\mu}$ cannot simultaneously be large in those
cases. As shown in the left lower plot, a dominating and observable
cross section in the $\mu\tau$ channel arises for the Majorana
phases $\phi_1=\phi_2=\pi/2$. One may check that in this case $h$
takes a special form in which the $h_{ee}$ and $h_{\mu\tau}$ are
large and the other entries small. This corresponds to the
approximate conservation of the $L_\mu - L_\tau$ flavor symmetry
\cite{Choubey:2004hn}.

We conclude from Fig.~\ref{fig:cross} that the di-muon channel looks
most promising among the possible final states, and therefore study
it further. The upper bounds of the cross section
$\sigma(e^-e^-\rightarrow \mu^-\mu^-)$ with respect to
$\sin\theta_{13}$ and $\delta$ are illustrated in
Fig.~\ref{fig:contour}. One reads from the plots that sizable cross
sections could be expected in the case $\delta \sim n\pi$ or
$\theta_{13} \sim 0$ if the neutrino mass spectrum is nearly
degenerate. Furthermore, even if the light neutrino mass spectrum is
hierarchical, there are still certain parameter regions, i.e.,
$(\delta,\sin\theta_{13}) \sim (\pi,0.04)$ for the normal mass
hierarchy and $(\delta,\sin\theta_{13}) \sim (0,0.01)$ for the
inverted mass hierarchy, that allow for identifying the Higgs
triplet since the expected number of events would be greater than 50
for an integrated luminosity $80~{\rm fb}^{-1}$. This is in
agreement with our the analytical expressions. For example, in the
normal mass ordering case, for $\delta=\pi$, Eq.~\eqref{eq:normal}
yields $\sin\theta_{13} \simeq 0.04$. In what regards the inverted
hierarchy, in the limit $m_3\sim 0$ the leading order contributions
in $h_{e\mu}$ disappear for $\phi_1=\phi_2=0$ as shown in
Eq.~\eqref{eq:inverted}. Thus, one should take into account the
``next-to-leading order'' contributions, which is given by
\begin{eqnarray}
\left|h_{e\mu}\right|^2 \simeq \frac{m^2_2}{4v^2_{\rm L}} \left(
\varepsilon^2 s^2_{12}c^2_{12} - 2 \varepsilon s_{12} c_{12}
s_{13}\cos\delta + s^2_{13} \right) , \label{eq:invert-hem}
\end{eqnarray}
where we have expanded the formula based on small parameters
$\varepsilon=(m_2-m_1)/m_2\simeq $ and $s_{13}$. For $\cos\delta = 0\,
(2\pi)$, $h_{e\mu} =0$ in Eq.~\eqref{eq:invert-hem} requires roughly
$s_{13} \simeq s_{12}c_{12}(m_2-m_1)/m_2$, which corresponds to
$\sin\theta_{13} \simeq 0.007$. This result again confirms the
numerical analysis.

As discussed before, the cross section relies heavily on the
configuration of Majorana CP-violating phases. In particular, in
order to obtain observable signatures, one would expect that
$\phi_1$ and $\phi_2$ are almost degenerate. In
Fig.~\ref{fig:phase}, we show the upper bounds of the cross section
$\sigma(e^-e^-\rightarrow \mu^-\mu^-)$ in the $\phi_1-\phi_2$ plane
in the normal ordering case. The interesting parameter region is, as
expected, the diagonal line of $\phi_1-\phi_2 \simeq 0$.
Consequently, collider experiments may play a crucial role in
determining the Majorana CP-violating phases. Since $h$ is not
sensitive to neutrino mass ordering in the ND limit, e.g., $m_i
> 0.05 ~{\rm eV}$, similar parameter region for $\phi_1$ and
$\phi_2$ also exists in the inverted mass ordering case. If the mass
spectrum is hierarchical, it is difficult to acquire visible cross
sections unless some fine tuning exists as shown in
Eqs.~\eqref{eq:normal-hem} and \eqref{eq:invert-hem}.

%%%%%%%%%%%%%%%%%%%%%%%%%%%%%%%%%%%%%%%%%
%%%%%%%%%%%%%%%%%%%% Fig.~5-6 %%%%%%%%%%%%%%%
\begin{figure}[t]
\begin{center}\vspace{-0.15cm}
\includegraphics[width=8cm]{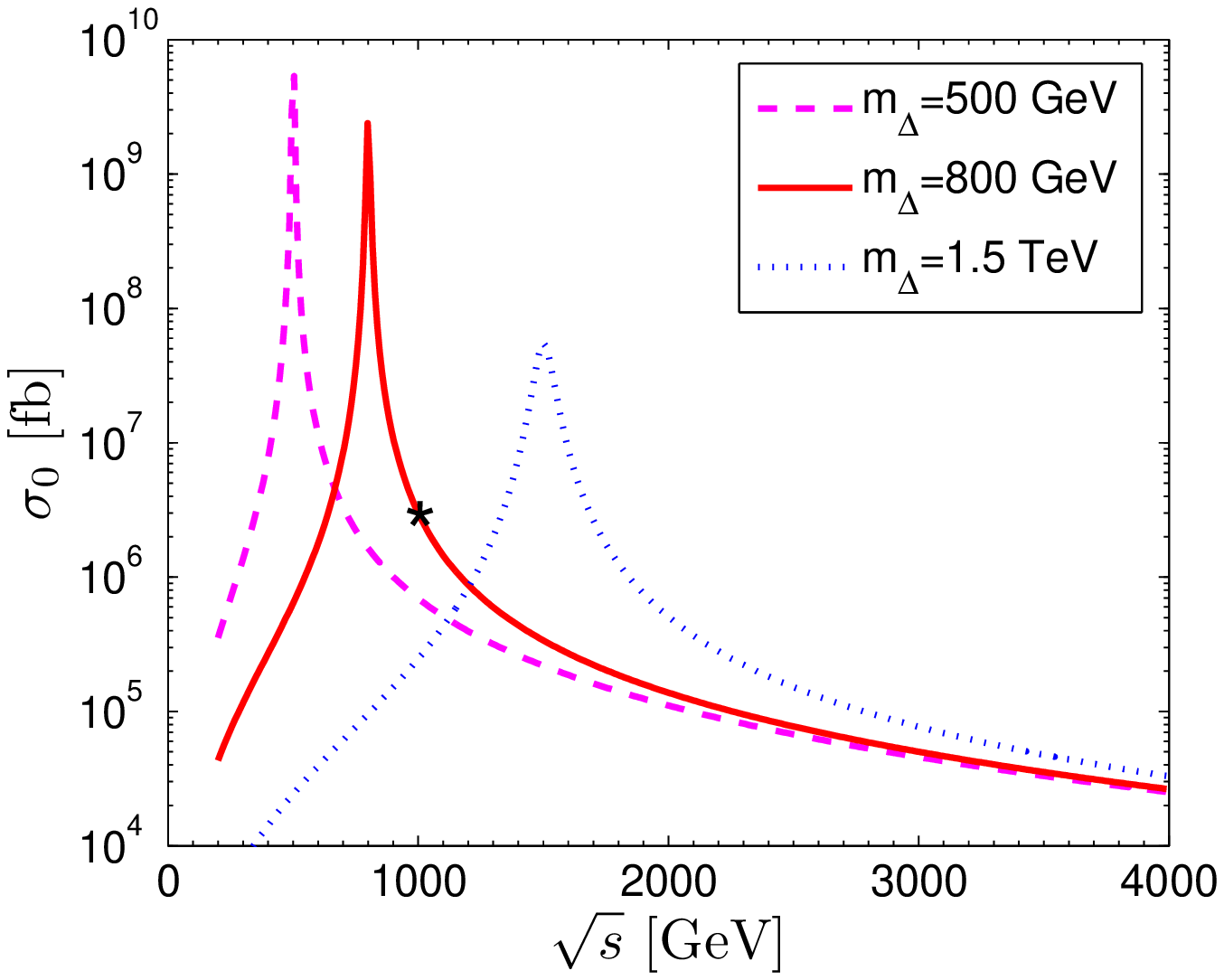}
\vspace{-0cm} \caption{\label{fig:s} The bare cross section
$\sigma_0$ (in units of fb) as a function of $\sqrt{s}$, with solid,
dashed and dotted lines corresponding to $m_\Delta=0.8~{\rm TeV}$,
$0.5~{\rm TeV}$ and $1.5~{\rm TeV}$, respectively. The black star on
the plot indicates the values of the parameters used in plotting
Figs.~(\ref{fig:cross}-\ref{fig:s}). For the definition of
$\sigma_0$, see Eq.~(\ref{eq:crossll}).}
\end{center}\vspace{0.4cm}
\includegraphics[width=8cm]{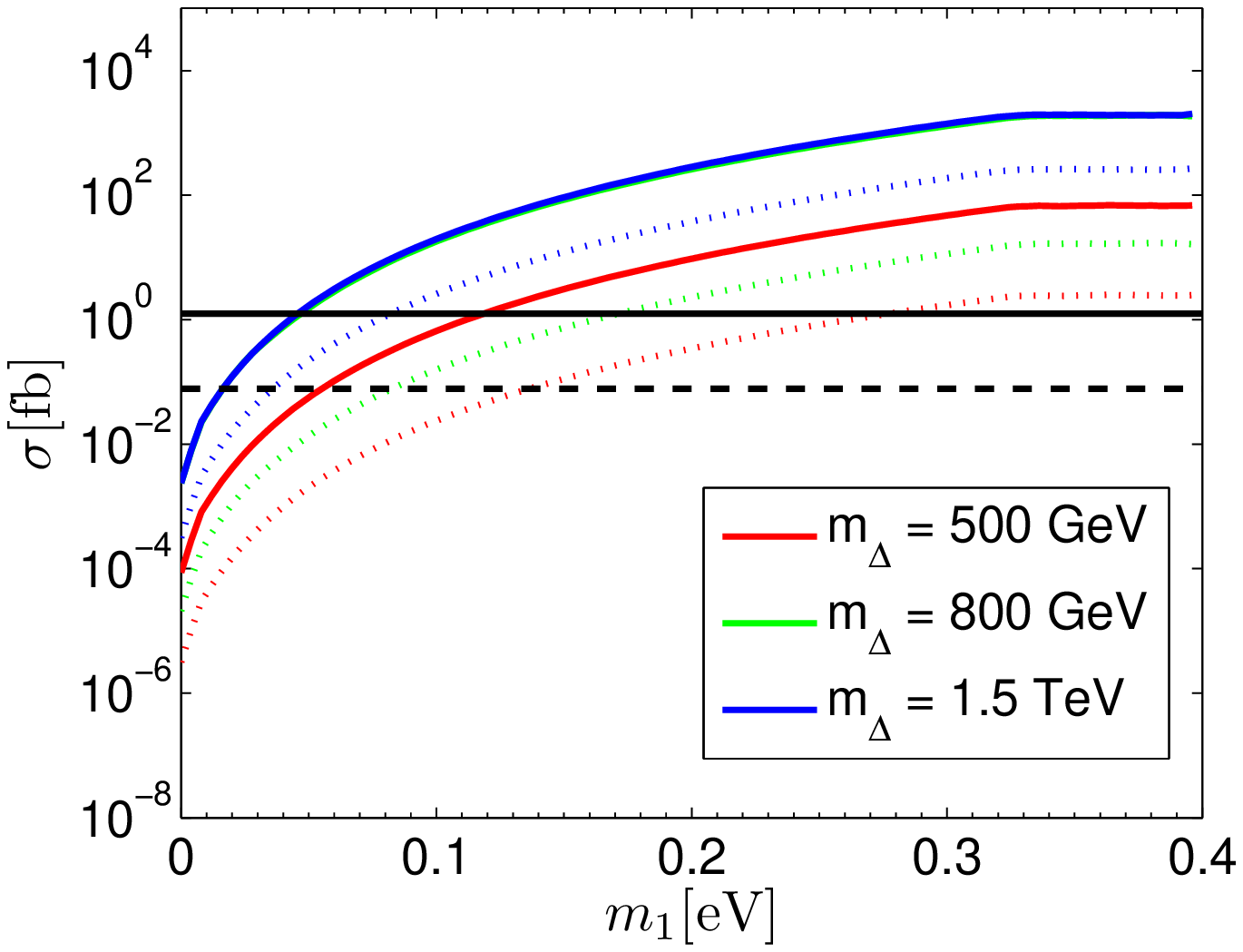}
\vspace{-0cm} \caption{\label{fig:6} Upper limits of the cross
section $e^-e^- \rightarrow \mu^-\mu^-$ with $\sqrt{s}=1~{\rm TeV}$
(solid lines) and  $\sqrt{s}=4~{\rm TeV}$ (dashed lines). The Higgs
triplet masses are labeled on the plot. Here we consider the normal
mass ordering case, and assume all the CP-violating phases and
$s_{13}$ to be zero for simplicity. The blue and green solid lines
for $\sqrt{s} = 1$ TeV and $m_\Delta = 800$ and $1500$ GeV are lying
basically on top of each other and are hardly distinguishable. 100
events correspond to $\sigma = 1.25~{\rm fb}$ (black solid line) and
$\sigma = 0.078~{\rm fb}$ (black dashed line), respectively.}
\end{figure}
%%%%%%%%%%%%%%%%%%%%%%%%%%%%%%%%%%%%%%%%%
%%%%%%%%%%%%%%%%%%%%%%%%%%%%%%%%%%%%%%%%%

Finally, one may wonder if the above estimate strongly depends on
the assumption on the mass of the doubly charged scalar, since
resonant production of $\Delta^{--}$ requires its mass to be known
with reasonable accuracy in order to tune the center-of-mass energy
of the colliding
electrons~\cite{Gunion:1995mq,Cuypers:1997qg,Raidal:1997tb}. In
Fig.~\ref{fig:s}, we give the bare cross section $\sigma_0$ [see
Eq.~(\ref{eq:crossll})] as a function of $\sqrt{s}$ for different
masses of the Higgs triplet. One reads from the plot that remarkable
cross sections can be naturally obtained without the need of a
severe fine-tuning of the colliding energy. To be concrete, we also
present in Fig.~\ref{fig:6} the upper limits of the cross section
$\sigma(e^-e^-\rightarrow \mu^-\mu^-)$ for the triplet masses
$m_\Delta = (0.5,0.8,1.5)~{\rm TeV}$ and $\sqrt{s}=(1,4)~{\rm TeV}$.
Recall that with the scaling behavior ${\cal L} = 80~(\sqrt{s}/{\rm
TeV})^2~{\rm fb}^{-1}$ one requires for 100 events cross sections of
$\sigma = 1.25~{\rm fb}$ and $\sigma = 0.078~{\rm fb}$, if $\sqrt{s}
= 1~{\rm TeV}$ and 4 TeV, respectively. Note that the cross sections
for $m_\Delta = 0.8~{\rm TeV}$ and $m_\Delta = 1.5~{\rm TeV}$ are
almost identical despite that $\sigma_0$ is different. This can be
understood from the LFV constraints on the Yukawa coupling $h$,
which becomes less severe for a larger triplet mass. This feature is
compensated by a reduced cross section. Recall that triplet masses
larger than 1~TeV cannot be probed at the LHC. Hence a future linear
collider running at the TeV scale can be a probe for such particles.
The bilepton channel at a linear collider is so spectacular that,
$\Delta^{--}$ with mass around TeV scale can be easily probed
together with much better sensitivity than that of the LHC.

\section{summary}
\label{sec:summary} We have studied the Higgs triplet mediated
processes $e^-e^- \rightarrow \alpha^- \beta^-$ at a future linear
collider run in a like-sign lepton mode. The strong dependence on
neutrino parameters and hence the flavor structure of the Majorana
mass matrix $m$ was emphasized and the di-muon channel was
identified as the most promising one. In order to avoid strong
constraints from lepton flavor violation, suppressing the $e\mu$
element of $m$ is required. The largest cross sections occur for a
near-diagonal mass matrix, which implies nearly degenerate neutrino
masses. However, even in the limit of a hierarchical neutrino
spectrum observable signatures could still be expected for certain
choices of neutrino mixing parameters. Therefore, measurements of
bilepton channels at a linear collider could be quite helpful in
order to provide valuable information on the neutrino parameters. In
addition, triplet masses beyond the reach of the LHC can be probed.

Finally, we would like to note that similar analyses could be
performed for a muon-collider (e.g., the process $\mu^- \mu^-
\rightarrow e^- e^-$), or an electron-muon facility, where in
addition interesting and characteristic processes like $e^- \mu^+
\rightarrow e^+ \mu^-$ could be studied.

\begin{acknowledgments}
This work was supported by the ERC under the Starting Grant MANITOP
and by the Deutsche Forschungsgemeinschaft in the Transregio 27
``Neutrinos and beyond -- weakly interacting particles in physics,
astrophysics and cosmology''. W.R.~wishes to thank Morimitsu Tanimoto
and the organizers of the Mini Workshop on Neutrinos at the IPMU, where
part of this paper was written.
\end{acknowledgments}

\bibliography{bib}

\end{document}